\documentclass[a4paper,12pt]{article}
\pdfoutput=1
\usepackage{multibib}
\usepackage{xpatch}
\usepackage{titling}
\usepackage{fontenc,fontaxes}
\usepackage{graphicx,booktabs}
\usepackage{mathtools}
\usepackage{amsmath,amsthm,amsfonts}
\usepackage{xspace}
\usepackage{authblk}
\usepackage[authoryear,round]{natbib}

\usepackage{comment}
\usepackage{setspace}
\usepackage[bookmarksnumbered,bookmarksopen,colorlinks,citecolor={blue}]{hyperref}
\usepackage{appendix}
\usepackage{tikz}
\usetikzlibrary{decorations.fractals,3d,positioning}
\usepackage{bbm}
\usepackage[bf,small]{caption}
\usepackage{doi}
\usepackage{subcaption}
\usepackage{pgfplots}
\usepackage[margin=1.25in]{geometry}
\usepackage[inline,shortlabels]{enumitem}
\setlist[enumerate,1]{label={\upshape(\roman*)}}

\mathtoolsset{mathic=true}

\theoremstyle{plain}
\newtheorem{thm}{Theorem}
\newtheorem*{thm*}{Theorem}
\newtheorem{lem}{Lemma}
\newtheorem*{lem*}{Lemma}
\newtheorem{prop}{Proposition}
\newtheorem*{prop*}{Proposition}

\newtheorem*{cor*}{Corollary}
\newtheorem{asmp}{Assumption}
\newtheorem*{asmp*}{Assumption}

\theoremstyle{definition}

\newtheorem*{defn*}{Definition}
\newtheorem{exmp}{Example}
\newtheorem*{exmp*}{Example}
\newtheorem{rem}{Remark}
\newtheorem*{rem*}{Remark}

\newtheoremstyle{step}{}{}{\itshape}{}{\itshape}{.}{.5em}{\thmname{#1} \thmnumber{#2}}
\theoremstyle{step}

\renewcommand\footnotemark{}
\renewcommand{\Pr}{\mathrm{P}}

\newcommand{\A}{\mathrm{A}}
\newcommand\E{\operatorname{E}}
\newcommand\diag{\operatorname{diag}}

\newcommand\supp{\operatorname{supp}}

\newcommand{\set}[1]{\left\{ #1 \right\} }

\newcommand{\e}{\mathrm{e}}
\newcommand{\diff}{\mathop{}\!\mathrm{d}}

\renewcommand{\Re}{\operatorname{Re}}
\renewcommand{\Im}{\operatorname{Im}}

\newcommand{\Z}{\mathbb{Z}}

\newcommand{\R}{\mathbb{R}}
\newcommand{\C}{\mathbb{C}}

\newcommand{\cI}{\mathcal{I}}
\newcommand{\cN}{\mathcal{N}}
\newcommand{\cS}{\mathcal{S}}

\newcommand{\eg}{\textrm{e.g.}\xspace}

\makeatletter
\newenvironment{mytitlepage}
{\begin{titlepage}\def\@thanks{}}
	{\end{titlepage}}
\xpatchcmd\titlepage{\setcounter{page}\@ne}{}{}{}
\xpatchcmd\endtitlepage{\setcounter{page}\@ne}{}{}{}
\makeatother

\begin{document}
	\begin{mytitlepage}
	\title{Determination of Pareto exponents in economic models driven by Markov multiplicative processes\thanks{We thank the Co-Editor and anonymous referees handling our submission, Andrew Atkeson, Alberto Bisin, Dan Cao, Xavier Gabaix, Fran\c{c}ois Geerolf, {\'E}milien Gouin-Bonenfant, Jim Hamilton, Gordon Hanson, Makoto Nirei, Taisuke Otsu, Werner Ploberger, Luciano Pomatto, John Stachurski, Omer Tamuz, Jonathan Weinstein, Lei Zhang, and participants at numerous conference and seminar presentations for helpful comments.}}
	
	\author[1]{Brendan K.\ Beare}
	\author[2]{Alexis Akira Toda}
	\affil[1]{School of Economics, University of Sydney}
	\affil[2]{Department of Economics, University of California San Diego}
		
	\maketitle
	\begin{center}
		Accepted for publication in \textit{Econometrica}.
	\end{center}	
	\bigskip	
	\begin{abstract}
		This article contains new tools for studying the shape of the stationary distribution of sizes in a dynamic economic system in which units experience random multiplicative shocks and are occasionally reset. Each unit has a Markov-switching type which influences their growth rate and reset probability. We show that the size distribution has a Pareto upper tail, with exponent equal to the unique positive solution to an equation involving the spectral radius of a certain matrix-valued function. Under a non-lattice condition on growth rates, an eigenvector associated with the Pareto exponent provides the distribution of types in the upper tail of the size distribution.
	\end{abstract}
	
	\end{mytitlepage}

\onehalfspacing

\section{Introduction}\label{intro}

This article presents new tools for studying the shape of the stationary distribution of a dynamic economic system. We have in mind models in which a population of economic units experience random multiplicative shocks to their size over time, occasionally perishing at random and being replaced with a new unit. An economic unit could be, for instance, a household with size measured by wealth, or a firm with size measured by market capitalization. Economic units have a type which may vary over time, such as a worker/entrepreneur type for households, or a productivity type for firms. The type affects the multiplicative growth rate, and perhaps also the survival rate, of an economic unit. Many heterogeneous-agent models of the Bewley-Aiyagari kind fit this general description if agents are subject to random mortality rather than having a certain finite or infinite lifespan.

It has long been known that random multiplicative growth is a generative mechanism for power laws. Random multiplicative growth plays a central role in economics and is known as Gibrat's law, while power laws are often referred to as Pareto tails. Early contributions to economics drawing a connection between random multiplicative growth and power laws include \cite{Champernowne1953}, \cite{WoldWhittle1957}, and \cite{SimonBonini1958}. The topic has attracted a resurgence of interest in economics following the publication of \cite{gabaix1999} and \citet{reed2001}; see \cite{gabaix2009,gabaix2016} and \cite{benhabibbisin2018} for partial surveys. In mathematics, the central contribution is \cite{kesten1973}. See \cite{Mitzenmacher2004} for a discussion of many other relevant contributions across a range of disciplines.

The primary contribution of this article is an equation whose unique positive solution is the Pareto (power law) exponent for the upper tail of the stationary distribution of sizes in a dynamic economic system of the kind described above. The equation is
\begin{equation}\label{eq:heuristicmain}
	\rho(\Pi\odot\Upsilon\odot\Psi(z))=1,
\end{equation}
where $\rho(\cdot)$ is the spectral radius (maximum modulus of eigenvalues) of a square matrix, $\odot$ is the Hadamard (entry-wise) product of matrices of the same size, $\Pi$ is a matrix of transition probabilities for types, $\Upsilon$ is a matrix of survival rates, and $\Psi(z)$ is a matrix of moment generating functions of random growth rates. If equation \eqref{eq:heuristicmain} admits a unique positive solution (which is the case under mild conditions), say $z=\alpha$, then our main result, Theorem \ref{thm:randomgrowth}, establishes that $\alpha$ is the Pareto exponent for the upper tail of the stationary distribution of sizes.

In the simple case where there is only a single type of economic unit (which, in the class of models we consider, implies serial independence of the multiplicative shocks to size and is therefore quite restrictive), equation \eqref{eq:heuristicmain} simplifies to
\begin{equation}\label{eq:manrubiazanette}
	\upsilon\psi(z)=1,
\end{equation}
where $\upsilon$ is the survival rate and $\psi(z)$ is the moment generating function of the random growth rate, both constant over time. Equation \eqref{eq:manrubiazanette} appears as Equation (10) in \citet{ManrubiaZanette1999}, and now seems to be fairly widely known. Applications include \cite{NireiAoki2016} and \cite{MukoyamaOsotimehin2019} in macroeconomics, \cite{MonteroVillarroel2013}, \cite{Yamamoto2014} and \cite{MeylahnSabhapanditTouchette2015} in statistical physics, and \citet{BeareToda2020PhysD} in epidemiology. Equation \eqref{eq:manrubiazanette} is also closely related to the Cram\'{e}r-Lundberg estimate of ruin probabilities used in actuarial science \cite[see e.g.][ch.~1]{EmbrechtsKluppelbergMikosch1997}.

Our equation \eqref{eq:heuristicmain} extends \eqref{eq:manrubiazanette} to a Markov setting in which economic units switch between types over time. There is a growing recognition in the literature that persistent heterogeneity in types plays a critical role in matching salient empirical regularities. \cite{benhabib-bisin-zhu2011} study an overlapping generations model in which agents are subject to random labor (additive) and capital (multiplicative) income shocks. They allow these shocks to be persistent through their dependence on a latent Markov state (i.e., type), writing (p.~127) ``the i.i.d.\ condition is very restrictive. Positive autocorrelations in [capital and labor income shocks] capture variations in social mobility in the economy, for example, economies in which returns on wealth and labor earning abilities are transmitted across generations.'' \cite{CaoLuo2017} study a similar model in which investors switch at random between high and low productivity types, while subject to random mortality. Regarding the importance of persistent heterogeneity in types, they write (p.~302) ``the model with homogeneous returns, when calibrated to match the salient aggregate statistics, produces a tail index that is an order of magnitude too high compared to the one in the data.'' Based on an analysis of Norwegian administrative tax records, \cite{FagerengGuisoMalacrinoPistaferri2020} confirm these observations on the importance of type dependence for explaining the upper tail of the wealth distribution, writing that (pp.~118--119) ``persistent traits of individual investors (such as financial sophistication, the ability to process and use financial information, the ability to overcome inertia, and—for entrepreneurs—the talent to manage and organize their businesses), are capable of generating persistent differences in returns to wealth that may be as relevant as those conventionally attributed in household finance to differences in risk exposure or scale''. Taking a somewhat different angle, \cite{GabaixLasryLionsMoll2016} emphasize the importance of persistent heterogeneity in productivity types not only for generating an income distribution with a tail sufficiently heavy to match U.S.\ data, but also for generating transition dynamics that are sufficiently fast to match the rate of increase of top income inequality. Additional articles discussing the importance of persistent heterogeneity in types for constructing plausibly calibrated models are discussed and cited therein (p.~2095).

Mathematically, the new results introduced in this article concern a class of Markov chains called \emph{Markov multiplicative processes with reset}, which we define in Section \ref{sec:randomgrowthprocess}. Loosely, these are processes driven by positive multiplicative shocks dependent on a Markov state variable, and subject to occasional reset to one at a state-dependent rate. After a heuristic discussion in Section \ref{sec:heuristic} aimed at building intuition, we state our main results in Section \ref{sec:randomgrowth}. Under mild conditions, these results establish \begin{enumerate*}
	\item that the left-hand side of equation \eqref{eq:heuristicmain} is convex in $z$ and admits at most one positive solution;
	\item a sufficient condition for the existence of a positive solution;
	\item the existence of a unique stationary distribution for a Markov multiplicative process with reset;
	\item a formula for the moment generating function of the stationary distribution (after taking logs);
	\item that a positive solution to equation \eqref{eq:heuristicmain} is the upper Pareto exponent for the stationary distribution.
\end{enumerate*}
In Section \ref{sec:lattice} we explore refinements obtaining under a non-lattice condition on growth rates, establishing that \begin{enumerate*}
	\item the upper tail of the stationary distribution satisfies a stronger form of Pareto decay;
	\item an eigenvector associated with the Pareto exponent provides the distribution of types in the upper tail.
\end{enumerate*}
We lay out some directions for future research in Section \ref{sec:conclusion}. A mathematical appendix to Sections \ref{sec:randomgrowth}--\ref{sec:lattice} (Appendix \ref{sec:proofmain}) contains the proofs of all numbered results, and a second appendix includes a discussion of comparative statics for the Pareto exponent (Appendix \ref{sec:compstatappx}).

Continuous-time versions of the results in this article have been established in a companion article, \citet{BeareSeoToda2021}. There we study the tail probabilities of a Markov-modulated L\'{e}vy process stopped at a state-dependent Poisson rate. The tail exponents in continuous-time are determined by an equation similar to \eqref{eq:heuristicmain}, but with a spectral abscissa taking the role of the spectral radius.

While we do not provide a substantive economic application of our results in this article, several recent articles and circulated manuscripts provide applications.\footnote{Articles in which our results are discussed sometimes cite an earlier version of this article titled ``Geometrically stopped Markovian random growth processes and Pareto tails'', available on the arXiv e-print repository since December 2017: \url{https://arxiv.org/abs/1712.01431}.} \citet[Thm.~4]{Toda2019} uses equation \eqref{eq:heuristicmain} to determine the Pareto exponent for the wealth distribution in a Huggett economy with stochastic discounting. \citet[Thm.~3.3]{MaStachurskiToda2020} do the same in a model with stochastic discounting, returns on wealth, and labor income. \citet[Prop.~3]{GomezGouinbonenfant2020} use our results to determine the Pareto exponent of wealth in an economy populated with entrepreneurs and rentiers. \citet[Prop.~A.10]{Gouinbonenfant2020} uses our results to calculate the Pareto exponent for the distribution of firm size in a model of the labor market, finding that Zipf's law is approximately satisfied. \citet{Gouin-BonenfantTodaParetoExtrapolation} show how to improve grid-based calculation of aggregate quantities in dynamic economic models by using equation \eqref{eq:heuristicmain} to extrapolate beyond the grid. \citet{BeareSeoToda2021} use a continuous-time version of equation \eqref{eq:heuristicmain} to study the tails of wealth in a Huggett economy inhabited by agents with constant absolute risk aversion.

\section{Markov multiplicative processes with reset}\label{sec:randomgrowthprocess}

Let $\Z_+$ be the set of nonnegative integers and $\cN=\set{1,\dots,N}$, a finite set. The elements of $\cN$ index the types of agents in a dynamic economic system. We will study the behavior of a homogeneous\footnote{Homogeneity means that $\Pr(S_{t+1}\leq s',J_{t+1}=n'\mid S_t=s,J_t=n)$ does not depend on $t$.} Markov chain $(S_t,J_t)_{t\in\Z_+}$ on $(0,\infty)\times\cN$. In each time period $t$, $S_t$ indicates the size of an agent (e.g.\ their wealth) and $J_t$ their type. The transition probabilities of our Markov chain are parametrized in terms of four objects.
\begin{enumerate}[(a)]
	\item \emph{Transition probabilities for $J_t$ without reset}: An $N\times N$ matrix $\Pi$ with nonnegative entries $\pi_{nn'}$, and all rows summing to one. The entries of $\Pi$ are the probabilities with which agents transition between types each period.
	\item \emph{Type-dependent survival probabilities}: An $N\times N$ matrix $\Upsilon$ whose entries $\upsilon_{nn'}$ are between zero and one inclusive. The entries of $\Upsilon$ are the survival probabilities for agents transitioning from type $n$ to type $n'$.
	\item \emph{Type probabilities for $J_t$ upon reset}: An $N\times 1$ vector $\varpi$ with nonnegative entries $\varpi_n$ summing to one. The entries of $\varpi$ provide the distribution of types for a new agent replacing an agent who has perished.
	\item \emph{Type-dependent growth rate distributions for $S_t$}: A collection of $N^2$ cumulative distribution functions (CDFs) on $(0,\infty)$, denoted $\Phi_{nn'}$ with $n,n'\in\cN$. The CDF $\Phi_{nn'}$ characterizes the distribution of growth rates for an agent transitioning from type $n$ to type $n'$.
\end{enumerate}
Conditional on having $S_t=s$ and $J_t=n$, the subsequent pair $(S_{t+1},J_{t+1})$ may be generated in the following way.
\begin{enumerate}
\item\label{item:SJ1} Choose $J_{t+1}^*$ at random from $\cN$, with $\Pr(J_{t+1}^*=n')=\pi_{nn'}$. This will be the value taken by $J_{t+1}$ contingent on reset not occurring.
\item\label{item:SJ2} Conditional on choosing $J_{t+1}^*=n'$ in Step \ref{item:SJ1}, reset occurs with probability $1-\upsilon_{nn'}$.
\item\label{item:SJ3} If reset occurs, then we set $S_{t+1}=1$ and choose $J_{t+1}$ at random from $\cN$, with $\Pr(J_{t+1}=n')=\varpi_{n'}$.
\item\label{item:SJ4} If reset does not occur, then we set $J_{t+1}=J_{t+1}^*$ and, conditional on choosing $J_{t+1}^*=n'$ in Step \ref{item:SJ1}, we draw a positive random variable $G_{nn'}$ from the CDF $\Phi_{nn'}$ and set $S_{t+1}=G_{nn'}s$.
\end{enumerate}
Steps \ref{item:SJ1}-\ref{item:SJ4} determine the Markov kernel for $(S_t,J_t)_{t\in\Z}$. Specifically, given $n,n'\in\cN$ and $s,s'>0$, the Markov kernel is
\begin{align}
	P_{sn}(s',n')&\coloneqq\Pr(S_{t+1}\leq s',J_{t+1}=n'\mid S_t=s,J_t=n)\notag\\
	&=\textcolor{gray}{\underbrace{\color{black}\pi_{nn'}\upsilon_{nn'}\Phi_{nn'}(s'/s)}_{J^*_{t+1}=n'\to\text{no reset}\to S_{t+1}\leq s'}}+\mathbbm{1}(s'\ge 1)\sum_{n''}\textcolor{gray}{\underbrace{\color{black}\pi_{nn''}(1-\upsilon_{nn''})\varpi_{n'}}_{J^*_{t+1}=n''\to\text{reset}\to J_{t+1}=n'}}.\label{eq:kernel}
\end{align}
We refer to a homogeneous Markov chain on $(0,\infty)\times\cN$ with Markov kernel given by \eqref{eq:kernel} as a \emph{Markov multiplicative process with reset}. We call the multiplicative factors $G_{nn'}$ gross growth rates, and their logarithms growth rates.

Under further regularity conditions, a Markov multiplicative process with reset can be made stationary with a suitable choice of its distribution at time $t=0$. This will be made clear in Proposition \ref{prop:stationary} below.

\section{Heuristic derivations}\label{sec:heuristic}

To build intuition, in this section we provide a brief heuristic derivation of our equation determining the Pareto exponent of a stationary Markov multiplicative process with reset. Let $(S_t,J_t)_{t\in\Z_+}$ be such a process, and conjecture that for all sufficiently large $s>1$ we have
\begin{equation}
\Pr(S_t>s,J_t=n)=y_ns^{-\alpha}\label{eq:jointtailprob}
\end{equation}
for some constant $y_n>0$ and Pareto exponent $\alpha>0$. Then
\begin{equation}
y_{n'}s^{-\alpha}=\Pr(S_{t+1}>s,J_{t+1}=n')=\sum_{n=1}^N\pi_{nn'}\upsilon_{nn'}\Pr(G_{nn'}S_t>s,J_t=n),\label{eq:heuristic1}
\end{equation}
where $G_{nn'}>0$ is the gross growth rate drawn from the CDF $\Phi_{nn'}$ in Step \ref{item:SJ4} in Section \ref{sec:randomgrowthprocess}. Applying the law of iterated expectations, we obtain
\begin{align}
	\Pr(G_{nn'}S_t>s,J_t=n)&=\E(\Pr(S_t>s/G_{nn'},J_t=n\mid G_{nn'}))\notag\\&=\E(y_n(s/G_{nn'})^{-\alpha})=y_n\E(G_{nn'}^\alpha)s^{-\alpha}.\label{eq:heuristic2}
\end{align}
For real $z$, set
\begin{equation}
\psi_{nn'}(z)\coloneqq\int_0^\infty s^z\diff\Phi_{nn'}(s),\label{eq:defM}
\end{equation}
the moment generating function (MGF) of $\log G_{nn'}$. Let $\Psi(z)$ be the $N\times N$ matrix-valued function with entries $\psi_{nn'}(z)$. Noting that $\psi_{nn'}(\alpha)=\E(G_{nn'}^\alpha)$, and combining \eqref{eq:heuristic1}, \eqref{eq:heuristic2} and \eqref{eq:defM}, we obtain
\begin{equation}
y_{n'}=\sum_{n=1}^Ny_n\pi_{nn'}\upsilon_{nn'}\psi_{nn'}(\alpha).\label{eq:heuristic3}
\end{equation}
Letting $y=(y_1,\dots,y_N)^\top$ and collecting \eqref{eq:heuristic3} into a vector, we obtain
\begin{equation*}
y^\top =y^\top(\Pi\odot\Upsilon \odot \Psi(\alpha)).
\end{equation*}
Thus $\Pi\odot\Upsilon\odot\Psi(\alpha)$ has a unit eigenvalue, with associated left eigenvector $y$.

The matrix $\Pi\odot\Upsilon\odot\Psi(\alpha)$ is nonnegative; suppose that it is also irreducible. The Perron-Frobenius theorem \citep[see e.g.][Thm.~8.4.4]{HornJohnson2013}, a fundamental result in the theory of Markov chains, asserts that the spectral radius of a square, nonnegative and irreducible matrix is the maximum real eigenvalue of that matrix, and that there are unique (up to positive scalar multiplication) left and right eigenvectors with positive entries corresponding to that eigenvalue. Since $y$ has positive entries and is a left eigenvector of $\Pi\odot\Upsilon \odot \Psi(\alpha)$ associated with a unit eigenvalue, we are thus led to suspect that $\Pi\odot\Upsilon\odot\Psi(\alpha)$ may have spectral radius equal to one. This suspicion, if valid, suggests the possibility that we may be able to determine the Pareto exponent $\alpha$ by finding a positive number $z$ that solves the equation \eqref{eq:heuristicmain}. The fact that this is indeed often possible is the main result of our article, Theorem \ref{thm:randomgrowth}.

The left eigenvector $y$ has a natural interpretation. It follows from \eqref{eq:jointtailprob} that
\begin{equation*}
\Pr(J_t=n \mid S_t>s)=\frac{\Pr(S_t>s,J_t=n)}{\Pr(S_t>s)}=\frac{y_ns^{-\alpha}}{\sum_{n=1}^Ny_ns^{-\alpha}}=\frac{y_n}{\sum_{n=1}^Ny_n}.
\end{equation*}
Therefore, if the left eigenvector $y$ is normalized such that its entries sum to one, then we have $\Pr(J_t=n \mid S_t>s)=y_n$. We may thus interpret $y$ to be, loosely, the distribution of types in the upper tail of the size distribution. We formalize this claim in Theorem \ref{thm:randomgrowthlattice}.

The preceding discussion is largely heuristic. The conjectured equality \eqref{eq:jointtailprob} cannot be expected to hold exactly except in very special cases. We will see that \eqref{eq:jointtailprob} holds approximately for large $s$ in fairly general settings. The above derivations based on \eqref{eq:jointtailprob} are therefore also approximate. We have skirted technical issues such as the existence of a unique stationary distribution for $(S_t,J_t)$, the finiteness of $\psi_{nn'}(\alpha)$, and the existence of a unique positive solution to \eqref{eq:heuristicmain}. Derivations similar to those above appear in \citet{Gouin-BonenfantTodaParetoExtrapolation}. We now turn to a rigorous statement of our main results.

\section{Main results}\label{sec:randomgrowth}

In this section we provide a rigorous statement of our primary result, which is that, under regularity conditions, a Markov multiplicative process with reset has a Pareto upper tail with exponent $\alpha$ given by the unique positive solution to equation \eqref{eq:heuristicmain}. We rely on the following regularity conditions.
\begin{asmp}\label{asmp:irreducible}
	\begin{enumerate*}
		\item\label{item:irreducible1} The matrix $\Pi\odot\Upsilon$ is irreducible.
		\item\label{item:irreducible2} There exist states $n,n'\in\cN$ such that $\pi_{nn'}>0$ and $\upsilon_{nn'}<1$.
		\item\label{item:irreducible3} $\psi_{nn'}\equiv 1$ whenever $\pi_{nn'}\upsilon_{nn'}=0$.
	\end{enumerate*}
\end{asmp}

Assumption \ref{asmp:irreducible}\ref{item:irreducible1} means that, for any pair of states $(n,n')$, if $J_t$ is in state $n$ then there is a positive probability of it eventually reaching state $n'$ before reset occurs. Assumption \ref{asmp:irreducible}\ref{item:irreducible2} means that there is some state $n$ such that, if $J_t$ is in state $n$, then reset occurs next period with positive probability. These two conditions together ensure that $J_t$ visits all states infinitely often and that reset occurs infinitely often. Assumption \ref{asmp:irreducible}\ref{item:irreducible3}, imposed without loss of generality, is merely a normalization since the growth rate when $J_t$ transitions from state $n$ to state $n'$ is unidentified if this transition never occurs.

To clarify the heuristic discussion in Section \ref{sec:heuristic}, the first thing we will do is restrict the domain of the matrix-valued function $\Psi(z)$ such that each of its entries $\psi_{nn'}(z)$, defined in \eqref{eq:defM}, is finite-valued. Define the set
\begin{equation*}
	\cI=\set{z\in \R:\text{$\psi_{nn'}(z)<\infty$ for all $n,n'\in \cN$}}.
\end{equation*}
Since MGFs are always convex and are equal to one at zero, $\cI$ is a convex set containing zero. We restrict the domain of $\Psi(z)$ to $\cI$ and, for $z\in\cI$, define the $N\times N$ matrix-valued function
\begin{equation}
\A(z)=\Pi\odot\Upsilon\odot\Psi(z). \label{eq:defA}
\end{equation}
Using \eqref{eq:defA}, the equation \eqref{eq:heuristicmain} may now be rewritten more simply as $\rho(\A(z))=1$.

We claimed in Section \ref{sec:heuristic} that it is often the case that there is a unique positive value of $z$ that solves the equation $\rho(\A(z))=1$. The following result concerning the shape of the function $\rho(\A(z))$ helps to explain why this is true.

\begin{prop}\label{prop:lambdaconv}
	The spectral radius $\rho(\A(z))$ is a convex function of $z\in\cI$ and satisfies $\rho(\A(0))\leq1$. If Assumption \ref{asmp:irreducible} holds, then $\rho(\A(0))<1$ and the equation $\rho(\A(z))=1$ has at most one positive solution $z=\alpha\in\cI$ and at most one negative solution $z=-\beta\in\cI$.
\end{prop}
Figure \ref{fig:alphabeta} depicts a typical shape for $\rho(\A(z))$ as a function of $z\in\cI$. The graph of $\rho(\A(z))$ is convex, less than one at zero (note that $\A(0)=\Pi\odot\Upsilon$), and diverging to infinity at the left and right endpoints of $\cI$ (which may in general be finite or infinite). There is a unique positive number $\alpha$ at which $\rho(\A(z))$ crosses one. We will see that $\alpha$ is the Pareto exponent for the upper tail of the stationary distribution of $S_t$. There is also a unique negative number $-\beta$ at which $\rho(\A(z))$ crosses one. We will see that $\beta$ is the Pareto exponent for the lower tail of the stationary distribution of $S_t$.

\begin{figure}
	\centering
	\begin{tikzpicture}[x=0.5cm,y=0.5cm,z=0.3cm,scale=2]
		\draw[thick, ->] (xyz cs:x=-2.7) -- (xyz cs:x=3) node[below] {\small $z$};
		\draw[thick, ->] (xyz cs:y=0) -- (xyz cs:y=3) node[right] {\small $\rho(\A(z))$};
		\node[below] at (xyz cs:x=-1.28) {\small $-\beta$};
		\node[below] at (xyz cs:x=.94) {\small $\alpha$};
		\node[below] at (xyz cs:x=0) {\small $0$};
		\node[above left] at (xyz cs:y=1) {\small $1$};
		\draw[smooth,samples=400,variable=\x,domain={-1.769}:{1.955}] plot (\x,{-.3+e^(.5*(\x/2))/(1-((1/1.3)-(1/1))*(\x/2)-(1/(1*1.3))*(\x/2)*(\x/2))},0);
		\draw[dashed] (xyz cs:x=-2.3,y=0) -- (xyz cs:x=-2.3,y=3);
		\draw[dashed] (xyz cs:x=2.6,y=0) -- (xyz cs:x=2.6,y=3);
		\draw[dashed] (xyz cs:x=-1.22,y=1) -- (xyz cs:x=.88,y=1);
		\draw[dashed] (xyz cs:x=-1.22,y=0) -- (xyz cs:x=-1.22,y=1);
		\draw[dashed] (xyz cs:x=.88,y=0) -- (xyz cs:x=.88,y=1);
		\filldraw (xyz cs:x=-2.3) circle (.6pt);
		\filldraw (xyz cs:x=2.6) circle (.6pt);
		\draw[<-] (xyz cs:x=-2.3,y=-.1) -- (xyz cs:x=-2.3,y=-.5);
		\node[below right,font=\small, align=left] at (xyz cs:x=-3.7,y=-.5) {\small Left endpoint of $\cI$};
		\draw[<-] (xyz cs:x=2.6,y=-.1) -- (xyz cs:x=2.6,y=-.5);
		\node[below left,font=\small, align=right] at (xyz cs:x=4,y=-.5) {\small Right endpoint of $\cI$};
	\end{tikzpicture}
	\caption{Determination of $\alpha$ and $\beta$ from the spectral radius of $\A(z)$.}\label{fig:alphabeta}
\end{figure}

The graph of $\rho(\A(z))$ does not always have the typical shape depicted in Figure \ref{fig:alphabeta}. Three atypical cases in which no positive $z$ solves $\rho(\A(z))=1$ are depicted in Figure \ref{fig:nonexistence}. If all growth rates are nonpositive, so that $\Pr(G_{nn'}\leq1)=1$ for all $n,n'\in\cN$, then $\rho(\A(z))$ is nonincreasing in $z$ and we have the case depicted in Figure \ref{fig:nonexistenceA}. If any growth rate does not have a light upper tail, so that the corresponding MGF $\psi_{nn'}(z)$ is infinite for all positive $z$, then the right endpoint of $\cI$ is zero and we have the case depicted in Figure \ref{fig:nonexistenceB}. Problems may also arise if the tails of growth rates are insufficiently light. For instance, suppose for simplicity that we have no Markov modulation ($N=1$) and that $\log G$ has PDF
\begin{equation*}
	f(x)=\begin{cases}c(x+2)^{-2}\e^{-ax}&\text{for }x\geq-1\\0&\text{for }x<-1,\end{cases}
\end{equation*}
where $a>0$ and $c$ is a positive constant such that the PDF integrates to one. In this case $\E(G^z)$ is finite for $z\leq a$ and infinite for $z>a$, so that $\cI=(-\infty,a]$. The existence of a positive solution to $\rho(\A(z))=1$ thus depends on whether $\rho(\A(a))$ is greater than one. The case where $a=1$ and $\upsilon=0.7$ is depicted in Figure \ref{fig:nonexistenceC}; we see that $\rho(\A(a))<1$, so that there is no positive solution to $\rho(\A(z))=1$.

The following result establishes that, to have a unique positive solution to $\rho(\A(z))=1$, it suffices that all growth rates have exponential moments of all orders, with at least one growth rate $\log G_{nn}$ positive with positive probability. An analogous result applies symmetrically to the existence of a unique negative solution to $\rho(\A(z))=1$.
\begin{prop}\label{prop:existence}
	Suppose that Assumption \ref{asmp:irreducible} is satisfied. If $\E(G_{nn'}^z)<\infty$ for all $z>0$ and all $n,n'\in\cN$, and if $\Pr(G_{nn}>1)>0$ for some $n\in\cN$, then there is a unique positive value of $z$ in the interior of $\cI$ such that $\rho(\A(z))=1$.
\end{prop}

\begin{figure}
	\centering
	\begin{subfigure}{0.32\linewidth}
		\centering
		\begin{tikzpicture}[x=0.5cm,y=0.5cm,z=0.3cm,scale=1.5]
			\draw[thick, ->] (xyz cs:x=-2.7) -- (xyz cs:x=3) node[below] {\scriptsize $z$};
			\draw[thick, ->] (xyz cs:y=0) -- (xyz cs:y=3) node[right] {\scriptsize $\rho(\A(z))$};
			\node[below] at (xyz cs:x=-.6) {\scriptsize $-\beta$};
			\node[below] at (xyz cs:x=0) {\scriptsize $0$};
			\node[above left] at (xyz cs:y=1) {\scriptsize $1$};
			\draw[smooth,samples=400,variable=\x,domain={-1.45}:{2.7}] plot (\x,{.7*(1/(1+.5*\x))},0);
			\draw[dashed] (xyz cs:x=-.6,y=1) -- (xyz cs:x=2.7,y=1);
			\draw[dashed] (xyz cs:x=-.6,y=0) -- (xyz cs:x=-.6,y=1);
			\filldraw (xyz cs:x=-.6) circle (.6pt);
		\end{tikzpicture}
		\caption{Nonpositive growth}\label{fig:nonexistenceA}
	\end{subfigure}
	\begin{subfigure}{0.32\linewidth}
		\centering
		\begin{tikzpicture}[x=0.5cm,y=0.5cm,z=0.3cm,scale=1.5]
			\draw[thick, ->] (xyz cs:x=-2.7) -- (xyz cs:x=3) node[below] {\scriptsize $z$};
			\draw[thick, ->] (xyz cs:y=0) -- (xyz cs:y=3) node[right] {\scriptsize $\rho(\A(z))$};
			\node[below] at (xyz cs:x=-1.4565) {\scriptsize $-\beta$};
			\node[below] at (xyz cs:x=0) {\scriptsize $0$};
			\node[above left] at (xyz cs:y=1) {\scriptsize $1$};
			\draw plot file {ParetoMGF.txt};
			\draw[dashed] (xyz cs:x=-1.4565,y=1) -- (xyz cs:x=2.7,y=1);
			\draw[dashed] (xyz cs:x=-1.4565,y=1) -- (xyz cs:x=-1.4565,y=0);
			\filldraw (xyz cs:x=-1.4565) circle (.6pt);
			\filldraw (xyz cs:y=0.7) circle (.6pt);
		\end{tikzpicture}
		\caption{Heavy-tailed growth}\label{fig:nonexistenceB}
	\end{subfigure}
	\begin{subfigure}{0.32\linewidth}
		\centering
		\begin{tikzpicture}[x=0.5cm,y=0.5cm,z=0.3cm,scale=1.5]
			\draw[thick, ->] (xyz cs:x=-2.7) -- (xyz cs:x=3) node[below] {\scriptsize $z$};
			\draw[thick, ->] (xyz cs:y=0) -- (xyz cs:y=3) node[right] {\scriptsize $\rho(\A(z))$};
			\node[on grid,below] at (xyz cs:x=-.496) {\scriptsize $-\beta$ \hspace{10 mm}};
			\node[below] at (xyz cs:x=2) {\scriptsize $a$ \hspace{10 mm}};
			\node[below] at (xyz cs:x=0) {\scriptsize $0$};
			\node[above left] at (xyz cs:y=1) {\scriptsize $1$};
			\draw plot file {EssentialMGF.txt};
			\draw[dashed] (xyz cs:x=-.496,y=1) -- (xyz cs:x=2.7,y=1);
			\draw[dashed] (xyz cs:x=-.496,y=1) -- (xyz cs:x=-.496,y=0);
			\draw[dashed] (xyz cs:x=2,y=0) -- (xyz cs:x=2,y=2.7);
			\filldraw (xyz cs:x=-.496) circle (.6pt);
			\filldraw (xyz cs:x=2,y=0.35) circle (.6pt);
		\end{tikzpicture}	
		\caption{Insufficiently light tail}\label{fig:nonexistenceC}
	\end{subfigure}
	\caption{Examples where no positive $z$ solves $\rho(\A(z))=1$.}\label{fig:nonexistence}
\end{figure}

We mentioned at the end of Section \ref{sec:randomgrowthprocess} that, under suitable regularity conditions, a Markov multiplicative process with reset can be made stationary with a suitable choice of its distribution at time $t=0$. The following result indicates that Assumption \ref{asmp:irreducible} provides sufficient regularity. It is proved by observing that reset generates a positive recurrent accessible atom of $(S_t,J_t)_{t\in\Z_+}$, which implies the existence of a unique stationary distribution.

\begin{prop}\label{prop:stationary}
	Let $(S_t,J_t)_{t\in\Z_+}$ be a Markov multiplicative process with reset. If Assumption \ref{asmp:irreducible} is satisfied, then there exists a unique probability distribution for $(S_0,J_0)$ such that $(S_t,J_t)_{t\in\Z_+}$ is stationary.
\end{prop}

We denote by $p$ the $N\times1$ vector of stationary probabilities for $J_t$, and by $q$ the $N\times1$ vector whose $n$th entry is $q_n=\sum_{n'}\pi_{nn'}(1-\upsilon_{nn'})$, the conditional probability of reset in period $t+1$ given $J_t=n$. The unconditional probability of reset is then $r\coloneqq\sum_np_nq_n$. It will also be useful to introduce notation for the subset of $\cI$ on which $\A(z)$ has spectral radius less than one. We therefore set $\cI_-=\set{z\in\cI:\rho(\A(z))<1}$, which is a convex subset of $\cI$ by Proposition \ref{prop:lambdaconv}, nonempty and containing zero under Assumption \ref{asmp:irreducible}. In typical cases such as the one depicted in Figure \ref{fig:alphabeta} we have $\cI_-=(-\beta,\alpha)$.

The following result, Proposition \ref{pro:Iminus}, is established in the course of proving our main result, Theorem \ref{thm:randomgrowth} below, but is also of independent interest. It reveals the form of the conditional MGF of $\log S_t$ given $J_t$, which has domain $\cI_-$. The notation $\mathrm{I}$ refers to an $N\times N$ identity matrix.
\begin{prop}\label{pro:Iminus}
	Let $(S_t,J_t)_{t\in\Z_+}$ be a stationary Markov multiplicative process with reset. Then, for each $z\in\cI_-$, the matrix $\mathrm{I}-\A(z)$ is invertible and
	\begin{equation*}
		\begin{bmatrix}
			\E(S_t^z\mathbbm{1}(J_t=1)) & \cdots & \E(S_t^z\mathbbm{1}(J_t=N))
		\end{bmatrix}
		=r\varpi^\top(\mathrm{I}-\A(z))^{-1}.
	\end{equation*}
\end{prop}
An immediate consequence of Proposition \ref{pro:Iminus} is that the MGF of $\log S_t$ is given by $\E(S_t^z)=r\varpi^\top(\mathrm{I}-\A(z))^{-1}1_N$ for $z\in\cI_-$, where $1_N$ is an $N\times1$ vector of ones. Proposition \ref{pro:Iminus} thus completely characterizes the stationary distribution of sizes whenever the interior of $\cI_-$ contains zero.

We now state our main result.
\begin{thm}\label{thm:randomgrowth}
	Let $(S_t,J_t)_{t\in\Z_+}$ be a stationary Markov multiplicative process with reset satisfying Assumption \ref{asmp:irreducible}. If the equation $\rho(\A(z))=1$ admits a unique positive solution $z=\alpha$ in the interior of $\cI$, then the limits inferior and superior of $s^\alpha\Pr(S_t>s)$ as $s\to\infty$ are positive and finite, and
	\begin{equation}\label{eq:weakPareto}
		\lim_{s\to\infty}\frac{\log\Pr(S_t>s)}{\log s}=-\alpha.
	\end{equation}
\end{thm}

Equation \eqref{eq:weakPareto} indicates that there is an approximately linear relationship in log-log scale between the tail probability $\Pr(S_t>s)$ and the threshold $s$ for large $s$, with slope $-\alpha$. In this sense, the upper tail of the distribution of $S_t$ is Pareto with exponent $\alpha$. The fact that this approximately linear relationship in log-log scale arises in distributions with a Pareto upper tail is the basis for the popular log-log rank-size regression method of estimating the Pareto exponent; see \eg \citet{GabaixIbragimov2011}. Note that Theorem \ref{thm:randomgrowth} does not assert the convergence of $s^\alpha\Pr(S_t>s)$ to a positive and finite limit, which would be a stronger notion of Pareto tail decay. This stronger property is not satisfied in general, but may be guaranteed by imposing a non-lattice condition on the distribution of growth rates, as discussed in Section \ref{sec:lattice}. To obtain the convergence \eqref{eq:weakPareto}, it suffices that $s^\alpha\Pr(S_t>s)$ has positive and finite limits inferior and superior. To see why, note that the latter condition implies the existence of positive and finite constants $c_-$ and $c_+$ such that $c_-\leq s^\alpha\Pr(S_t>s)\leq c_+$ for all sufficiently large $s$. Taking logarithms, dividing by $\log s$, and letting $s\to\infty$, we obtain \eqref{eq:weakPareto}.

We close this section with four remarks on Theorem \ref{thm:randomgrowth} and a brief example.

\begin{rem}\label{rem:lowertail}
By replacing $S_t$ with $1/S_t$, it is easy to deduce from Theorem \ref{thm:randomgrowth} that the lower tail probability $s^{-\beta}\Pr(S_t<s)=(1/s)^{\beta}\Pr(1/S_t>1/s)$ has a positive and finite limit as $s\downarrow 0$, where $z=-\beta$ is the unique \emph{negative} solution to the equation $\rho(\A(z))=1$ in the interior of $\cI$ (if it exists). Thus the stationary distribution of $S_t$ also exhibits a power law in the \emph{lower} tail.\footnote{Although not widely known, lower tail power law behavior in the distribution of economic aggregates has been documented in a number of empirical studies. For city size, see the bottom-right panel of Figure 1 of \citet{reed2001} (who writes ``lower-tail power-law behaviour\ldots is not apparently widely recognized''), top panels of Figures 1 and 2 of \citet{reed2002} (who writes ``the lower-tail plots exhibit linearity''), Footnote 8 of \citet{giesen-zimmermann-suedekum2010}, who write ``Among the 100 smallest cities we also find a distinctive power law pattern'', and Figure 2 of \citet{DevadossLucksteadDanforthAkhundjanov2016}. Aside from city size, power law behavior in the lower tail has been documented in income (bottom-left panel of Figure 1 of \citealp{reed2001}; Figure 3 of \citealp{Toda2011PRE}; Figure 1 of \citealp{Toda2012JEBO}) and consumption (Figure 2 of \citealp{TodaWalsh2015JPE}; and \citealp{Toda2017MD}).}
\end{rem}

\begin{rem}\label{rem:homog}
If the survival probabilities $\upsilon_{nn'}$ are constant across states, so that we may write $\upsilon_{nn'}=\upsilon\in (0,1)$, then the equation $\rho(\A(z))=1$ reduces to $\upsilon\rho(\Pi \odot\Psi(z))=1$. If in addition the growth rate MGFs $\psi_{nn'}$ depend only on the current state $n'$, so that we may write $\psi_{nn'}=\psi_{n'}$, then letting $\mathrm{D}(z)=\diag(\psi_1(z),\dots,\psi_N(z))$, the equation $\rho(\A(z))=1$ reduces to
\begin{equation}
\upsilon\rho(\Pi\mathrm{D}(z))=1.\label{eq:homogM}
\end{equation}
Similarly, if the growth rate MGFs $\psi_{nn'}$ depend only on the \emph{previous} state $n$, then the equation $\rho(\A(z))=1$ reduces to $\upsilon\rho(\mathrm{D}(z)\Pi)=1$, which is identical to \eqref{eq:homogM} noting that $\rho(AB)=\rho(BA)$ in general.
\end{rem}

\begin{rem}\label{rem:compstat}
	Our characterization of the upper Pareto exponent $\alpha$ in Theorem \ref{thm:randomgrowth} is implicit, in the sense that it is given by the positive solution to $\rho(\A(z))=1$. It may be useful for economic applications to provide comparative statics for $\alpha$; that is, results indicating how $\alpha$ varies as we vary the parameters that enter into the matrix $\A(z)$. We provide a rigorous statement of such results in Appendix \ref{sec:compstatappx}. The main findings are as follows.
\begin{enumerate*}
	\item A marginal increase in any survival rate $\upsilon_{nn'}$ reduces $\alpha$.
	\item A marginal increase in the mean of any growth rate $\log G_{nn'}$ reduces $\alpha$.
	\item A marginal increase in the scale of any growth rate $\log G_{nn'}$ reduces $\alpha$.
	\item If the survival probabilities $\upsilon_{nn'}$ and growth rates $\log G_{nn'}$ depend only on the current state $n'$, or only on the previous state $n$, then a marginal increase in the persistence of the Markov modulator $J_t$ reduces $\alpha$.
\end{enumerate*}
\end{rem}

\begin{rem}
	Proposition \ref{pro:Iminus} and Theorem \ref{thm:randomgrowth} are related to Theorem 15 in \citet{Toda2014JET} on geometric sums. The latter result establishes that the Laplace distribution provides a small $p$ asymptotic approximation to the distribution of a sum of $T_p$ weakly dependent random variables, where $T_p$ is a geometric random variable with success probability $p$. It implies a double Pareto approximation to the stationary distribution of $S_t$ when the survival probability does not vary with types and is close to one. Proposition \ref{pro:Iminus} provides the exact MGF of $\log S_t$, while Theorem \ref{thm:randomgrowth} shows that the tail probabilities of $S_t$ decay at Pareto rates determined by the positive and negative solutions to $\rho(\A(z))=1$, with neither result relying on asymptotic approximation or type-invariance of survival rates.
\end{rem}

\begin{exmp}\label{exmp:twostate}
	Suppose that there are two states ($N=2$). By the Perron-Frobenius theorem, $\rho(\A(z))$ is the maximum real eigenvalue of $\A(z)$, and so by applying the usual formula for the eigenvalues of a $2\times2$ matrix \citep[see e.g.][p.~39]{HornJohnson2013} we compute
	\begin{equation}
		\rho(\A(z))=\frac{1}{2}\left(a_{11}(z)+a_{22}(z)+\sqrt{\left(a_{11}(z)-a_{22}(z)\right)^2+4a_{12}(z)a_{21}(z)}\right),\label{eq:rhotwostate}
	\end{equation}
	where $a_{nn'}(z)\coloneqq\pi_{nn'}\upsilon_{nn'}\psi_{nn'}(z)$. Setting $\rho(\A(z))$ equal to one and solving for $z$ gives a unique positive solution $z=\alpha$ if the conditions in Proposition \ref{prop:existence} are satisfied. To illustrate concretely, suppose that
	\begin{equation*}
		\Pi=\begin{bmatrix}1-\pi_{12}&\pi_{12}\\\pi_{12}&1-\pi_{12}\end{bmatrix}
	\end{equation*}
	for some $\pi_{12}\in(0,1)$, that $\upsilon_{nn'}=\upsilon=.95$, and that each $\psi_{nn'}$ is Gaussian, with locations $\mu_{11}=\mu_{21}=.03$ and $\mu_{12}=\mu_{22}=.01$ and scales $\sigma_{nn'}=\sigma=.01$. We graph the unique positive solution $z=\alpha$ to $\rho(\A(z))=1$ as a function of $\pi_{12}$ in Figure \ref{fig:numericalA}. The graph shows that the Pareto exponent $\alpha$ increases smoothly from around 1.7 to around 2.56 as the transition probability $\pi_{12}$ increases from zero to one. The fact that $\alpha$ is increasing in $\pi_{12}$ is consistent with the comparative statics presented in Appendix \ref{sec:compstatappx} and previewed in Remark \ref{rem:compstat}. The lower limit of approximately 1.7 is the value of $z$ solving the equation $\upsilon\psi_{11}(z)=1$; i.e.\ it is the Pareto exponent for an agent always of the higher growth type. The upper limit of approximately 2.56 is the value of $z$ solving the equation $\upsilon\psi(z)=1$, where $\psi$ is an equally weighted mixture of $\psi_{12}$ and $\psi_{21}$. We will discuss Figure \ref{fig:numericalB} when we return to this example in Section \ref{sec:lattice}.
	
	\begin{figure}
		\captionsetup[subfigure]{justification=centering}
		\centering
		\begin{subfigure}{0.4\linewidth}
			\centering
			\begin{tikzpicture}[x=0.5cm,y=0.5cm,z=0.3cm,scale=6]
				\draw[thick, -] (xyz cs:x=0) -- (xyz cs:x=1) node[below] {\scriptsize $1$};
				\draw[thick, ->] (xyz cs:y=0) -- (xyz cs:y=1);
				\node[below] at (xyz cs:x=0) {\scriptsize $0$};
				\node[left] at (xyz cs:x=0) {\scriptsize $1.5$};
				\node[left] at (xyz cs:y={(1.7-1.5)/1.5}) {\scriptsize $1.7$};
				\node[left] at (xyz cs:y={(2.56-1.5)/1.5}) {\scriptsize $2.56$};
				\node[left] at (xyz cs:y=1) {\scriptsize $3$};
				\draw plot file {alpha.txt};
				\draw[dashed] (xyz cs:x=0,y={(2.56-1.5)/1.5}) -- (xyz cs:x=1,y={(2.56-1.5)/1.5});
				\filldraw (xyz cs:y={(1.7-1.5)/1.5}) circle (.15pt);
				\node at (xyz cs:x=-.3,y=.5) {\footnotesize $\alpha$};
				\node at (xyz cs:x=.5,y=-.2) {\footnotesize $\pi_{12}$};
			\end{tikzpicture}
			\caption{Upper Pareto exponent}\label{fig:numericalA}
		\end{subfigure}
		\begin{subfigure}{0.4\linewidth}
			\centering
			\begin{tikzpicture}[x=0.5cm,y=0.5cm,z=0.3cm,scale=6]
				\draw[thick, -] (xyz cs:x=0) -- (xyz cs:x=1) node[below] {\scriptsize $1$};
				\draw[thick, -] (xyz cs:y=0) -- (xyz cs:y=1) node[left] {\scriptsize $1$};
				\node[below left] at (xyz cs:x=0) {\scriptsize $0$};
				\node[left] at (xyz cs:y=0.5) {\scriptsize $0.5$};
				\draw plot file {typeprobs.txt};
				\draw[dashed] (xyz cs:x=0,y=.5) -- (xyz cs:x=1,y=.5);
				\filldraw (xyz cs:y=1) circle (.15pt);
				\node at (xyz cs:x=-.3,y=.5) {\footnotesize $y_1$};
				\node at (xyz cs:x=.5,y=-.2) {\footnotesize $\pi_{12}$};
			\end{tikzpicture}
			\caption{Type 1 share in upper tail}\label{fig:numericalB}
		\end{subfigure}
		\caption{Upper Pareto exponent and type 1 share in Example \ref{exmp:twostate}.}\label{fig:numerical}
	\end{figure}
\end{exmp}

Examples 3.5 and 3.6 in \citet{BeareSeoToda2021} concern a continuous-time reformulation of Example \ref{exmp:twostate} in which the log-size of agents evolves as a two-state Brownian motion with drift. There, the determination of the Pareto exponent boils down to examining the roots of a quartic polynomial, or of a quadratic polynomial in the absence of a diffusive component. The latter case corresponds closely to the economic model in \citet{CaoLuo2017}, in which the upper Pareto exponent for the stationary distribution of wealth is given by the unique positive root of a quadratic polynomial.

\section{Refinements under a non-lattice condition}\label{sec:lattice}

While Theorem \ref{thm:randomgrowth} establishes Pareto decay of the upper tail of the stationary distribution of $S_t$ in the sense of there being an asymptotically linear relationship in log-log scale between the tail probability $\Pr(S_t>s)$ and threshold $s$ with slope $-\alpha$, it is notable that Theorem \ref{thm:randomgrowth} does not assert the convergence of $s^\alpha\Pr(S_t>s)$ to a positive and finite limit. Such convergence does not hold in general. We show in this section that it obtains under a non-lattice condition on growth rates. We also establish a characterization of upper-tail type shares that obtains under the non-lattice condition.

\begin{asmp}\label{asmp:lattice}
	There do not exist constants $d>0$ and $c_{nn'}$, $n\neq n'$, such that
	\begin{enumerate*}
		\item\label{asmp:lattice1} $\mathrm{supp}(\log G_{nn})\subset d\Z$ for each $n\in\cN$, and
		\item\label{asmp:lattice2} $\mathrm{supp}(\log G_{nn'})\subset c_{nn'}+d\Z$ for each $n,n'\in\cN$ with $n\neq n'$.
	\end{enumerate*}
\end{asmp}

A simple sufficient (but not necessary) condition for Assumption \ref{asmp:lattice} is that at least one growth rate is not a discrete random variable. The following result strengthens the conclusion of Theorem \ref{thm:randomgrowth} when Assumption \ref{asmp:lattice} is satisfied, and also shows that in this case a left eigenvector of $\A(\alpha)$ characterizes type shares in the upper tail of $S_t$.

\begin{thm}\label{thm:randomgrowthlattice}
	Let $(S_t,J_t)_{t\in\Z_+}$ be a stationary Markov multiplicative process with reset satisfying Assumptions \ref{asmp:irreducible} and \ref{asmp:lattice}. If the equation $\rho(\A(z))=1$ admits a unique positive solution $z=\alpha$ in the interior of $\cI$, then $s^\alpha\Pr(S_t>s)$ converges to a positive and finite limit as $s\to\infty$, so that $S_t$ has a Pareto upper tail with exponent $\alpha$. Moreover, $\A(\alpha)$ has a unique left eigenvector $y=(y_1,\dots,y_N)^\top$ with strictly positive entries summing to one, and for each $n\in\cN$ we have
	\begin{equation}\label{eq:eigenvectorconvergencelattice}
		\lim_{s\to\infty}\Pr(J_t=n\mid S_t>s)=y_n.	
	\end{equation}
\end{thm}

We illustrate the statement about limiting type shares in Theorem \ref{thm:randomgrowthlattice} by revisiting Example \ref{exmp:twostate}.
{
\renewcommand{\theexmp}{1}
\begin{exmp}[continued]
	At $z=\alpha$, a left eigenvector with strictly positive entries associated with the maximum real eigenvalue computed in \eqref{eq:rhotwostate} is
	\begin{equation*}
		\tilde{y}=\begin{bmatrix}\frac{1}{2}\left(a_{11}(\alpha)-a_{22}(\alpha)+\sqrt{\left(a_{11}(\alpha)-a_{22}(\alpha)\right)^2+4a_{12}(\alpha)a_{21}(\alpha)}\right)\\a_{12}(\alpha)\end{bmatrix}.
	\end{equation*}
	Scaling the entries of $\tilde{y}$ such that they sum to one yields the left eigenvector $y$ referred to in Theorem \ref{thm:randomgrowthlattice}. In Figure \ref{fig:numericalB}, adopting the conditionally Gaussian parametrization used in our earlier discussion of this example, we graph $y_1$, the share of type 1 agents in the upper tail of the size distribution, as a function of the transition probability $\pi_{12}$. The graph shows that the share of type 1 agents in the upper tail decreases smoothly from one to one half as the transition probability $\pi_{12}$ increases from zero to one.
\end{exmp}
\addtocounter{exmp}{-1}
}

The following simple example shows that the conclusions of Theorem \ref{thm:randomgrowthlattice} need not be satisfied if Assumption \ref{asmp:lattice} is dropped.

\begin{exmp}\label{exmp:lattice}
	Suppose again that $N=2$, and let $(S_t,J_t)$ be a Markov multiplicative process with reset on $(0,\infty)\times\{1,2\}$ with $G_{11}=G_{22}=1$, $G_{12}=4$ and $G_{21}=2$, and with survival rate $\upsilon\in(0,1)$ constant across states. Suppose that we set $S_t=1$ and $J_t=1$ upon reset, and that in the absence of reset $J_t$ has transition probability matrix
	\begin{equation*}
		\Pi=\begin{bmatrix}0&1\\1&0\end{bmatrix}.
	\end{equation*}
	These assumptions fully specify the transition probabilities of $(S_t,J_t)$. Noting that Assumption \ref{asmp:irreducible} is satisfied, we deduce from Proposition \ref{prop:stationary} that $(S_t,J_t)$ has a unique stationary distribution, and assume it to be initialized at this distribution.
	
	Assumption \ref{asmp:lattice} is not satisfied in this example, as may be seen by taking $d=1$ and $c_{12}=2c_{21}=2\log 2$. We show in Appendix \ref{sec:proofmain} that the equation $\rho(\A(z))=1$ admits a unique positive solution $z=\alpha=-(2/3)\log_2\upsilon$, and that
	\begin{align*}
		\upsilon=\liminf_{s\to\infty}s^{\alpha}\Pr(S_t>s)&<\limsup_{s\to\infty}s^{\alpha}\Pr(S_t>s)=\upsilon^{-1/3},\quad\text{and}\\
		\frac{\upsilon}{1+\upsilon}=\liminf_{s\to\infty}\Pr(J_t=1\mid S_t>s)&<\limsup_{s\to\infty}\Pr(J_t=1\mid S_t>s)=\frac{1}{1+\upsilon}.
	\end{align*}
	The conclusions of Theorem \ref{thm:randomgrowthlattice} therefore do not hold. Nevertheless, Theorem \ref{thm:randomgrowth} implies that
	\begin{align*}
		\lim_{s\to\infty}\frac{\log\Pr(S_t>s)}{\log s}=-\alpha=\frac{2}{3}\log_2\upsilon.
	\end{align*}
\end{exmp}

Example \ref{exmp:lattice} illustrates the fragility of Theorem \ref{thm:randomgrowthlattice}. Since any collection of growth rate distributions satisfying Assumption \ref{asmp:lattice} can be well-approximated by a collection of growth rate distributions not satisfying it, and vice-versa, it is natural to be skeptical of results which depend on this condition for their validity. Non-lattice conditions have been used elsewhere to establish Pareto tail behavior of stationary solutions to stochastic difference equations.\footnote{See, for instance, (1.11) in Theorem A of \citet{kesten1973}, (2) in Theorems 1 and 2 of \citet{desaporta2005}, (A7) in Assumption 1.2 of \citet{roitershtein2007}, and the ``spread out'' condition on pp.\ 1408--9 of \citet{collamore2009}; and in an economic application, see Footnote 56 in \cite{benhabib-bisin-zhu2011}. Note also that results of this kind typically exclude the possibility of reset: see (1.9) in Theorem A of \citet{kesten1973}, the requirement that zero be excluded from the state space in Theorems 1 and 2 of \citet{desaporta2005}, (A5) in Assumption 1.2 of \citet{roitershtein2007}, and the requirement that $\log A_n$ be finite-valued on p.\ 1408 of \citet{collamore2009}.} An advantage of our approach is that our primary result, Theorem \ref{thm:randomgrowth}, does not rely on any non-lattice condition for its validity. The form of Pareto tail decay given in \eqref{eq:weakPareto} is in this sense robust.

\section{Final remarks}\label{sec:conclusion}

Given the importance of persistent heterogeneity in productivity types for constructing plausibly calibrated economic models, we expect to see modellers adopting this feature more widely in future, and hope that the results we have presented here may prove useful to them. We conclude by outlining three potential generalizations of our results which may broaden the scope of applications.

First, it may be useful to pursue a relaxation of our assumption of irreducibility. The evolution of types may follow a reducible Markov chain if, for instance, the health of agents randomly and irreversibly deteriorates over time, affecting their access to different productivity types. We rely on irreducibility at various points in our technical arguments. In particular, in the proof of Theorem \ref{thm:appxresult} in Appendix \ref{sec:proofmain}, which is used to prove Theorems \ref{thm:randomgrowth} and \ref{thm:randomgrowthlattice}, irreducibility is critical to establishing that poles in the MGF of log-size determining the Pareto exponents are simple. The reducible case may require a more general treatment of these poles.

Second, it may be useful to pursue a relaxation of our assumption that the number of types is finite. This would allow, for instance, productivity to be modelled as a general autoregressive process. Such a generalization would likely entail adapting our numerous arguments involving matrices such that they apply with linear operators on infinite dimensional spaces.

Third, it may be useful to generalize our results such that they apply when the random growth of agents is only asymptotically, rather than exactly, multiplicative. For instance, while capital income is naturally associated with multiplicative growth in wealth, the effect of labor income on wealth is additive. In an economy in which agents randomly accrue both capital and labor income, the growth in the wealth of the wealthiest agents is predominantly driven by capital income, and thus approximately multiplicative. Since it is the wealthiest agents which determine the upper Pareto exponent, we expect that our results on the upper Pareto exponent may be applied in a setting of this sort. This has been done in \citet{Gouin-BonenfantTodaParetoExtrapolation}, with heuristic justification. A rigorous demonstration of the validity of our characterization of the upper Pareto exponent under asymptotically multiplicative growth would place such applications on firmer footing.

\appendix

\numberwithin{thm}{section}
\numberwithin{table}{section}
\numberwithin{exmp}{section}
\numberwithin{prop}{section}
\numberwithin{defn}{section}
\numberwithin{lem}{section}
\numberwithin{rem}{section}
\numberwithin{cor}{section}
\numberwithin{equation}{section}

\section{Proofs of claims in Sections \ref{sec:randomgrowth} and \ref{sec:lattice}}\label{sec:proofmain}

\begin{proof}[Proof of Proposition \ref{prop:lambdaconv}]
Since each entry of $\A(z)$ is a nonnegative multiple of a MGF finite-valued on $\cI$, a result of \citet{Kingman1961} guarantees that the spectral radius $\rho(\A(z))$ is a convex function of $z\in\cI$. And since $\A(0)$ is nonnegative and bounded entry-wise by $\Pi$, which is nonnegative with row sums of one, Theorems 8.1.22 and 8.4.5 in \citet{HornJohnson2013} imply that $\rho(\A(0))\leq\rho(\Pi)=1$. Under Assumption \ref{asmp:irreducible}, $\A(0)$ is nonnegative and irreducible with at least one entry strictly less than the corresponding entry of $\Pi$, so Problem 15 on p.~515 in \citet{HornJohnson2013} implies that $\rho(\A(0))<\rho(\Pi)$; thus in this case $\rho(\A(0))<1$. Since $\rho(\A(z))$ is convex on $\cI$, under Assumption \ref{asmp:irreducible} there can therefore be at most one positive and one negative solution to the equation $\rho(\A(z))=1$ in $\cI$.
\end{proof}
\begin{proof}[Proof of Proposition \ref{prop:existence}]
	The assumption that $\E(G_{nn'}^z)<\infty$ for all $z>0$ and all $n,n'\in\cN$ guarantees that $[0,\infty)\subset\cI$ and that $\rho(\A(z))<\infty$ for all $z>0$, while the assumption that $\Pr(G_{nn}>1)>0$ guarantees that $\psi_{nn}(z)\to\infty$ as $z\to\infty$. We therefore have
	\begin{align*}
		\rho(\A(z))\geq\rho(\diag(\A(z)))=\max\{\pi_{nn}\upsilon_{nn}\psi_{nn}(z):n\in\cN\}\to\infty
	\end{align*}
	as $z\to\infty$, using a monotonicity property of the spectral radius \citep[Thm.~8.1.18]{HornJohnson2013} to obtain the inequality, and noting that $\pi_{nn}\upsilon_{nn}>0$ under Assumption \ref{asmp:irreducible}\ref{item:irreducible3}. Under Assumption \ref{asmp:irreducible}, Proposition \ref{prop:lambdaconv} implies that $\rho(\A(0))<1$ and that $\rho(\A(z))$ is convex, hence continuous, on the interior of $\cI$. It follows from the intermediate value theorem that $\rho(\A(z))=1$ for some positive $z$ in the interior of $\cI$.
\end{proof}

\begin{proof}[Proof of Proposition \ref{prop:stationary}]
	It will be convenient to augment the state space $\cN$ such that we are able to keep track of when reset occurs. To this end, let $\widetilde{\cN}=\set{\widetilde{n}_1,\dots,\widetilde{n}_{2N}}$, where
	\begin{equation*}
		\widetilde{n}_i=\begin{cases}(i,0)&\text{for }i=1,\dots,N,\\
			(i-N,1)&\text{for }i=N+1,\dots,2N.
			\end{cases}
	\end{equation*}
    Let $K_t$ be equal to one if there is reset at time $t$, or equal to zero otherwise. The sequence of pairs $(J_t,K_t)_{t\in\Z_+}$ is then a Markov chain on $\widetilde{\cN}$. Fix $n\in\cN$ with $\varpi_n>0$. Under Assumption \ref{asmp:irreducible}, since $\varpi_n>0$, the state $(n,1)$ is accessible from any other state; therefore, since there are finitely many states, the state $(n,1)$ must be positive recurrent \citep[Cor.~7.2.3]{DoucMoulinesPriouretSoulier2018}.
    
    Consider the singleton $\mathcal R_n\coloneqq\set{1}\times\set{n}$. Since $(S_t,J_t)\in\mathcal R_n$ whenever $(J_t,K_t)=(n,1)$, the fact that $(n,1)$ is a positive recurrent accessible state of $(J_t,K_t)_{t\in\Z_+}$ implies that $\mathcal R_n$ is a positive recurrent accessible atom of $(S_t,J_t)_{t\in\Z_+}$. It therefore follows from Theorem 6.4.2 in \citet{DoucMoulinesPriouretSoulier2018} that there exists a unique probability distribution for $(S_0,J_0)$ such that $(S_t,J_t)_{t\in\Z_+}$ is stationary.
\end{proof}

Until now we have regarded $\A(z)$ to be a function of a real variable $z\in\cI$, where $\cI$ was defined to be the smallest convex subset of the real line on which all of the MGFs $\psi_{nn'}(z)$ are finite. In what follows, it will be more useful to regard $\A(z)$ and each of the MGFs $\psi_{nn'}(z)$ to be functions of a complex variable. We therefore extend $\cI$ to a strip in the complex plane by setting $\cS=\set{z\in\C:\Re(z)\in\cI}$. Owing to the fact that
\begin{equation}\label{eq:triangle}\int |s^z|\diff\Phi_{nn'}(s)=\int s^{\Re(z)}\diff\Phi_{nn'}(s)=\psi_{nn'}(\Re(z)),
\end{equation}
it is assured that each $\psi_{nn'}(z)\coloneqq\int s^z\diff\Phi_{nn'}(s)$ is well-defined as a complex-valued function of $z\in\cS$, and that $\A(z)$ is well-defined as a complex matrix-valued function of $z\in\cS$. The functions $\psi_{nn'}(z)$ are called Mellin-Stieltjes transforms of the corresponding CDFs $\Phi_{nn'}(s)$. We also extend $\cI_-$ to a strip in the complex plane by setting $\cS_-=\set{z\in\C:\Re(z)\in\cI_-}$. Figure \ref{fig:strips} illustrates $\cS$ and $\cS_-$.

\renewcommand{\thefigure}{A.1}
\begin{figure}
	\centering
	\begin{tikzpicture}[x=0.5cm,y=0.5cm,z=0.3cm,scale=1.9]
		\draw[thick, ->] (xyz cs:x=-2.6) -- (xyz cs:x=3) node[below] {\scriptsize $\Re(z)$};
		\draw[thick, ->] (xyz cs:y=0) -- (xyz cs:y=2) node[right] {\scriptsize $\rho(\A(z))$};
		\draw[thick, ->] (xyz cs:z=-2) -- (xyz cs:y=.1,z=2) node[left] {\scriptsize $\Im(z)$};
		\node[below] at (xyz cs:x=-1.28) {\scriptsize $-\beta$};
		\node[below] at (xyz cs:x=.94) {\scriptsize $\alpha$};
		\node[below] at (xyz cs:x=0) {\scriptsize $0$};
		\node[above left] at (xyz cs:x=.1,y={2/3-.1}) {\scriptsize $1$};
		\draw[smooth,samples=400,variable=\x,domain={-1.769}:{1.955}] plot (\x,{(2/3)*(-.3+e^(.5*(\x/2))/(1-((1/1.3)-(1/1))*(\x/2)-(1/(1*1.3))*(\x/2)*(\x/2)))},0);
		\draw[dashed] (xyz cs:x=-2.3,y=0) -- (xyz cs:x=-2.3,y=2);
		\draw[dashed] (xyz cs:x=2.6,y=0) -- (xyz cs:x=2.6,y=2);
		\draw[dashed] (xyz cs:x=-1.22,y={2/3}) -- (xyz cs:x=.88,y={2/3});
		\draw[dashed] (xyz cs:x=-1.22,y=0) -- (xyz cs:x=-1.22,y={2/3});
		\draw[dashed] (xyz cs:x=.88,y=0) -- (xyz cs:x=.88,y={2/3});
		\draw[thick,<->] (xyz cs:x=-2.3,z=-2) -- (xyz cs:x=-2.3,z=2);
		\draw[thick,<->] (xyz cs:x=2.6,z=-2) -- (xyz cs:x=2.6,z=2);
		\begin{scope}[canvas is xz plane at y=0,transform shape]
			\fill[opacity=.2] (-2.3,1.9) rectangle (2.6,-1.9);
		\end{scope}
		\node[font=\small,align=left] at (xyz cs:x=1.5,y=0,z=-1.5) {$\cS$};
	\end{tikzpicture}
	\begin{tikzpicture}[x=0.5cm,y=0.5cm,z=0.3cm,scale=1.9]
		\draw[thick, ->] (xyz cs:x=-2.6) -- (xyz cs:x=3) node[below] {\scriptsize $\Re(z)$};
		\draw[thick, ->] (xyz cs:y=0) -- (xyz cs:y=2) node[right] {\scriptsize $\rho(\A(z))$};
		\draw[thick, ->] (xyz cs:z=-2) -- (xyz cs:y=.1,z=2) node[left] {\scriptsize $\Im(z)$};
		\node[below] at (xyz cs:x=-1.28) {\scriptsize $-\beta$};
		\node[below] at (xyz cs:x=.94) {\scriptsize $\alpha$};
		\node[below] at (xyz cs:x=0) {\scriptsize $0$};
		\node[above left] at (xyz cs:x=.1,y={2/3-.1}) {\scriptsize $1$};
		\draw[smooth,samples=400,variable=\x,domain={-1.769}:{1.955}] plot (\x,{(2/3)*(-.3+e^(.5*(\x/2))/(1-((1/1.3)-(1/1))*(\x/2)-(1/(1*1.3))*(\x/2)*(\x/2)))},0);
		\draw[dashed] (xyz cs:x=-2.3,y=0) -- (xyz cs:x=-2.3,y=2);
		\draw[dashed] (xyz cs:x=2.6,y=0) -- (xyz cs:x=2.6,y=2);
		\draw[dashed] (xyz cs:x=-1.22,y={2/3}) -- (xyz cs:x=.88,y={2/3});
		\draw[dashed] (xyz cs:x=-1.22,y=0) -- (xyz cs:x=-1.22,y={2/3});
		\draw[dashed] (xyz cs:x=.88,y=0) -- (xyz cs:x=.88,y={2/3});
		\draw[thick,<->] (xyz cs:x=-2.3,z=-2) -- (xyz cs:x=-2.3,z=2);
		\draw[thick,<->] (xyz cs:x=2.6,z=-2) -- (xyz cs:x=2.6,z=2);
		\draw[thick,<->] (xyz cs:x=-1.28,z=-2) -- (xyz cs:x=-1.28,z=2);
		\draw[thick,<->] (xyz cs:x=.94,z=-2) -- (xyz cs:x=.94,z=2);
		\begin{scope}[canvas is xz plane at y=0,transform shape]
			\fill[opacity=.2] (-1.28,1.9) rectangle (.94,-1.9);
		\end{scope}
		\node[font=\small,align=left] at (xyz cs:x=.38,y=0,z=-1.5) {$\cS_-$};
	\end{tikzpicture}
	\caption{The strips $\cS$ and $\cS_-$ in the complex plane.}\label{fig:strips}
\end{figure}

\begin{lem}\label{lem:Sminus}
	For each $z\in\cS_-$ we have $\rho(\A(z))<1$.
\end{lem}
\begin{proof}[Proof of Lemma \ref{lem:Sminus}]
	In view of \eqref{eq:triangle}, the entries of $\A(z)$ are bounded in modulus by the corresponding entries of $\A(\Re(z))$ for all $z\in\cS$. Using a monotonicity property of the spectral radius \citep[Thm.~8.1.18]{HornJohnson2013} we therefore have $\rho(\A(z))\leq\rho(\A(\Re(z)))$ for all $z\in\cS$. The desired result now follows from the definitions of $\cI_-$ and $\cS_-$.
\end{proof}
The quantity $\E(S_t^z)$, viewed as a function of the complex variable $z$, is called the Mellin transform of the distribution of $S_t$. The following lemma provides a formula for $\E(S_t^z\mathbbm{1}(J_t=n))$ valid for each $z\in\cS_-$.
\begin{lem}\label{lem:Mellinformula}
	Let $(S_t,J_t)_{t\in\Z_+}$ be a stationary Markov chain on $(0,\infty)\times\cN$ with Markov kernel given by \eqref{eq:kernel}. Then, for each $z\in\cS_-$, the matrix $\mathrm{I}-\A(z)$ is invertible and
	\begin{equation}\label{eq:ESz}
		\begin{bmatrix}
		\E(S_t^z\mathbbm{1}(J_t=1)) & \cdots & \E(S_t^z\mathbbm{1}(J_t=N))
		\end{bmatrix}
		=r\varpi^\top(\mathrm{I}-\A(z))^{-1}.
	\end{equation}
\end{lem}
\begin{proof}[Proof of Lemma \ref{lem:Mellinformula}]
	Let $F_n(s)\coloneqq\Pr(S_t\leq s,J_t=n)$. Then, using \eqref{eq:kernel},
	\begin{align}
        &F_{n'}(s')=\sum_n\int\Pr(S_{t+1}\leq s',J_{t+1}=n'\mid S_t=s,J_t=n)\diff F_n(s)\notag\\
        &=\sum_n\pi_{nn'}\upsilon_{nn'}\int\Phi_{nn'}(s'/s)\diff F_n(s)+\mathbbm{1}(s'\ge 1)\varpi_{n'}\sum_{n,n''}p_n\pi_{nn''}(1-\upsilon_{nn''}).
		\label{eq:conv}
    \end{align}
Noting that $\int (s')^z\Phi_{nn'}(\diff s'/s)=s^z\psi_{nn'}(z)$ for each $z\in\cS$, and using the definition of the unconditional reset probability $r$, we take the Mellin-Stieltjes transform of either side of \eqref{eq:conv} to obtain
\begin{equation}
	\int s^z\diff F_{n'}(s)
	=\sum_n \pi_{nn'}\upsilon_{nn'}\psi_{nn'}(z)\int s^z\diff F_n(s)+r\varpi_{n'}.\label{eq:mellin}
\end{equation}	
Letting $L(z)$ denote the $N\times1$ vector with $n$th entry
\begin{equation*}
\E(S_t^z\mathbbm{1}(J_t=n))=\int s^z\diff F_n(s),
\end{equation*}
we may rewrite \eqref{eq:mellin} using matrix notation as $L(z)=\A(z)^\top L(z)+r\varpi$ or, equivalently,
\begin{equation}\label{eq:MellinS}
	L(z)^\top(\mathrm{I}-\A(z))=r\varpi^\top.
\end{equation}
Equation \eqref{eq:MellinS} is valid for each $z\in\cS$. Lemma \ref{lem:Sminus} implies that the matrix $\mathrm{I}-\A(z)$ is invertible for each $z\in\cS_-$. We may therefore postmultiply both sides of \eqref{eq:MellinS} by $(\mathrm{I}-\A(z))^{-1}$ to obtain \eqref{eq:ESz}, valid for each $z\in\cS_-$.
\end{proof}
\begin{proof}[Proof of Proposition \ref{pro:Iminus}]
	Immediate from Lemma \ref{lem:Mellinformula}.
\end{proof}

\begin{thm}\label{thm:appxresult}
	Suppose that Assumption \ref{asmp:irreducible} holds, and let $(S_t,J_t)_{t\in\Z_+}$ be a stationary homogeneous Markov chain on $(0,\infty)\times\cN$ with Markov kernel given by \eqref{eq:kernel}. Suppose further that the equation $\rho(\A(z))=1$ admits a unique positive solution $z=\alpha$ in the interior of $\cI$. Then $\A(\alpha)$ has an algebraically simple unit eigenvalue, and the associated right and left eigenspaces are spanned by unique right and left eigenvectors $x$ and $y$ with positive entries summing to one. Let
	\begin{equation*}
		C=\frac{r\varpi^\top x}{y^\top \A'(\alpha)x},
		\label{eq:defC}
	\end{equation*}
	where $\A'(\alpha)$ is the matrix of complex derivatives of $\A(z)$ at $z=\alpha$. Let $B$ be the supremum of all $b>0$ such that $\mathrm{I}-\A(\alpha+i\tau)$ is invertible for $\tau\in(-b,b)$ except at $\tau=0$. Then $B\in(0,\infty]$, $C\in(0,\infty)$ and, for each $n\in\cN$,
	\begin{align}
		\frac{2\pi C/B}{\e^{2\pi\alpha/B}-1}y_n&\le \liminf_{s\to\infty}s^{\alpha}\Pr(S_t>s,J_t=n)\notag \\
		&\le \limsup_{s\to\infty}s^{\alpha}\Pr(S_t>s,J_t=n)\le \frac{2\pi C/B}{1-\e^{-2\pi\alpha/B}}y_n\label{eq:Wutail1}
	\end{align}
	if $B<\infty$, or otherwise
	\begin{equation}
		\lim_{s \to\infty}s^{\alpha}\Pr(S_t>s,J_t=n)=\frac{Cy_n}{\alpha}.\label{eq:Wutail2}
	\end{equation}
\end{thm}

\begin{proof}[Proof of Theorem \ref{thm:appxresult}]
	Let $\Omega$ denote the interior of $\cS$, an open and connected subset of the complex plane, nonempty since it contains $\alpha$. The matrix-valued function $\A(z)$ is holomorphic on $\Omega$. Since $\rho(\A(z))$ is a convex function of $z\in\cI$ with $\rho(\A(0))<1$ (Proposition \ref{prop:lambdaconv}), and $\rho(\A(\alpha))=1$, it must be the case that $\mathrm{I}-\A(z)$ is invertible at all real numbers between $0$ and $\alpha$, which are elements of $\Omega$.  It therefore follows from Theorem A.2 in \citet{BeareSeoToda2021} that $\mathrm{I}-\A(z)$ has a meromorphic inverse on $\Omega$, with poles at the points of noninvertibility of $\mathrm{I}-\A(z)$. Since $\A(\alpha)$ is nonnegative, we know from the Perron-Frobenius theorem that its spectral radius of one is an eigenvalue. Consequently, zero is an eigenvalue of $\mathrm{I}-\A(\alpha)$. We deduce that $\mathrm{I}-\A(z)$ is not invertible at $\alpha\in\Omega$, and that $\alpha$ is a pole of $(\mathrm{I}-\A(z))^{-1}$. Since every pole is an isolated singularity, it follows immediately that $B>0$.
	
	We now show that $\alpha$ is a simple pole of $(\mathrm{I}-\A(z))^{-1}$, and determine the associated residue. Since $\A(\alpha)$ is nonnegative and irreducible under Assumption \ref{asmp:irreducible}, it follows from the Perron-Frobenius theorem that the unit eigenvalue of $\A(\alpha)$ is algebraically simple, hence geometrically simple, and is associated with unique right and left eigenvectors $x,y$ with positive entries summing to one. From Theorem A.2 in \citet{BeareSeoToda2021} we thus deduce that $\alpha$ is a simple pole of $(\mathrm{I}-\A(z))^{-1}$, with residue
	\begin{align*}
		R &\coloneqq -x (y^\top\A'(\alpha)x)^{-1}y^\top = -cxy^\top,
	\end{align*}
	where $c\coloneqq(y^\top\A'(\alpha)x)^{-1}$ is a positive real number.
	
	Let $e^{(n)}$ denote an $N\times1$ vector with $n$th entry equal to one and other entries equal to zero. By Lemma \ref{lem:Mellinformula}, we have
	\begin{align}\label{eq:conditionalMellin}
		\E(S_t^z\mid J_t=n)&=(r/p_n)\varpi^\top(\mathrm{I}-\A(z))^{-1}e^{(n)}
	\end{align}	
	for $z\in\cS_-$, where $r$ is the unconditional probability of reset and $p_n=\Pr(J_t=n)$. Moreover, since $(\mathrm{I}-\A(z))^{-1}$ is meromorphic on $\Omega$, \eqref{eq:conditionalMellin} defines a meromorphic extension of $\E(S_t^z\mid J_t=n)$ to $\Omega$. As $z\to\alpha$, we obtain
	\begin{align*}
		(z-\alpha)\E(S_t^z\mid J_t=n)&=(r/p_n)\varpi^\top(z-\alpha)(\mathrm{I}-\A(z))^{-1}e^{(n)}\\
		&\to (r/p_n)\varpi^\top Re^{(n)}=-rc(\varpi^\top x)y_n/p_n=-Cy_n/p_n.
	\end{align*}
	The eigenvectors $x$ and $y$ have positive entries, and $\varpi$ has nonnegative entries with at least one entry positive, so $\varpi^\top x>0$ and $y_n>0$. The unconditional reset probability $r$ is positive under Assumption \ref{asmp:irreducible} and, as noted above, $c>0$, so also $C>0$. Therefore $\lim_{z\to\alpha}(z-\alpha)\E(S_t^z\mid J_t=n)\ne 0$. This shows that $\alpha$ is a simple pole of $\E(S_t^z\mid J_t=n)$ with residue $-Cy_n/p_n$. Letting $B'$ denote the supremum of all $b>0$ such that $\alpha$ is the unique singularity of $\E(S_t^z\mid J_t=n)$ on $\{\alpha+i\tau:\tau\in(-b,b)\}$, it now follows from Theorem A.1 in \citet{BeareSeoToda2021} that
	\begin{align}
		\frac{2\pi C/B'}{\e^{2\pi\alpha/B'}-1}y_n/p_n&\le \liminf_{s\to\infty}s^{\alpha}\Pr(S_t>s\mid J_t=n)\notag \\
		&\le \limsup_{s\to\infty}s^{\alpha}\Pr(S_t>s\mid J_t=n)\le \frac{2\pi C/B'}{1-\e^{-2\pi\alpha/B'}}y_n/p_n\label{eq:Wutail3}
	\end{align}
	if $B'<\infty$, or otherwise
	\begin{equation}
		\lim_{s \to\infty}s^{\alpha}\Pr(S_t>s\mid J_t=n)=\frac{Cy_n}{\alpha p_n}.\label{eq:Wutail4}
	\end{equation}
	It is apparent from \eqref{eq:conditionalMellin} that if $\alpha+i\tau$ is a singularity of $\E(S_t^z\mid J_t=n)$ then it is also a singularity of $(\mathrm{I}-\A(z))^{-1}$. Thus $B\leq B'$. If $B=\infty$ then also $B'=\infty$, and so \eqref{eq:Wutail2} follows from \eqref{eq:Wutail4}. If $B<\infty$ and $B'<\infty$ then \eqref{eq:Wutail1} follows from \eqref{eq:Wutail3} by noting that the lower (upper) bound in \eqref{eq:Wutail3} is increasing (decreasing) in $B'$. If $B<\infty$ and $B'=\infty$ then \eqref{eq:Wutail1} follows from \eqref{eq:Wutail4} by noting that $Cy_n/\alpha$ falls between the lower and upper bounds in \eqref{eq:Wutail1}.
	\end{proof}

	\begin{rem}
		Theorems A.1 and A.2 in \cite{BeareSeoToda2021}, which play a critical role in the proof of Theorem \ref{thm:appxresult}, are not novel to that paper but are stated there in a way which is convenient for our purposes. The role played by Theorem A.1 is particularly important. This result, due to \cite{Nakagawa2007}, is the source of the tail probability bounds in \eqref{eq:Wutail1}.
	\end{rem}

	\begin{proof}[Proof of Theorem \ref{thm:randomgrowth}]
	Theorem \ref{thm:appxresult} establishes that $s^\alpha\Pr(S_t>s)$ has positive and finite limits inferior and superior. There thus exist positive and finite constants $c_-$ and $c_+$ such that $c_-\leq s^\alpha\Pr(S_t>s)\leq c_+$ for all sufficiently large $s$. Taking logarithms, dividing by $\log s$, and letting $s\to\infty$, we obtain \eqref{eq:weakPareto}.
	\end{proof}
	\begin{lem}\label{lem:feller}
		Let $X$ be a real random variable and $\alpha$ a positive real number such that $\E(\e^{\alpha X})<\infty$, and let $\tau$ be a nonzero real number. Then $|\E(\e^{(\alpha+i\tau)X})|=\E(\e^{\alpha X})$ if and only if $\mathrm{supp}(X)\subset a+(2\pi/\tau)\Z$ for some $a\in\R$. Moreover, $\E(\e^{(\alpha+i\tau)X})=\E(\e^{\alpha X})$ if and only if $\mathrm{supp}(X)\subset (2\pi/\tau)\Z$.
	\end{lem}
    \begin{proof}[Proof of Lemma \ref{lem:feller}]
    	The result follows from an obvious modification to the proofs of Lemmas 3 and 4 in \citet[pp.~500--501]{feller1971}.
    \end{proof}

	\begin{proof}[Proof of Theorem \ref{thm:randomgrowthlattice}]
	The result follows from Theorem \ref{thm:appxresult} if we can show that $B=\infty$; that is, that $\alpha$ is the unique point of noninvertibility of $\mathrm{I}-\A(z)$ on the axis $\Re(z)=\alpha$. Suppose to the contrary that $\mathrm{I}-\A(\alpha+i\tau)$ is not invertible for some nonzero $\tau\in\R$. In view of \eqref{eq:triangle}, the entries of $\A(\alpha+i\tau)$ are bounded in modulus by the corresponding entries of $\A(\alpha)$. Using a monotonicity property of the spectral radius \citep[Thm.~8.1.18]{HornJohnson2013} we deduce that $\rho(\A(\alpha+i\tau))\leq\rho(\A(\alpha))=1$. But noninvertibility of $\mathrm{I}-\A(\alpha+i\tau)$ implies that $\A(\alpha+i\tau)$ has a unit eigenvalue, so we must have $\rho(\A(\alpha+i\tau))=\rho(\A(\alpha))=1$. It now follows from Theorem 8.4.5 in \citet{HornJohnson2013} that $\A(\alpha+i\tau)=\mathrm{D}\A(\alpha)\mathrm{D}^{-1}$, where $\mathrm{D}=\diag(\e^{i\theta_1},\dots,\e^{i\theta_N})$ for some $\theta_1,\dots,\theta_N\in\R$. We may therefore write
	\begin{align*}
		\pi_{nn'}\upsilon_{nn'}\psi_{nn'}(\alpha+i\tau)&=\pi_{nn'}\upsilon_{nn'}\e^{i(\theta_n-\theta_{n'})}\psi_{nn'}(\alpha)
	\end{align*}
	for all $n,n'\in\cN$. Consequently, $\psi_{nn}(\alpha+i\tau)=\psi_{nn}(\alpha)$ for all $n\in\cN$ such that $\pi_{nn}\upsilon_{nn}>0$, and $|\psi_{nn'}(\alpha+i\tau)|=\psi_{nn'}(\alpha)$ for all $n,n'\in\cN$ such that $\pi_{nn'}\upsilon_{nn'}>0$. Moreover, for any $n,n'\in\cN$ such that $\pi_{nn'}\upsilon_{nn'}=0$, we trivially have $\psi_{nn'}(\alpha+i\tau)=\psi_{nn'}(\alpha)=1$ due to the normalization in Assumption \ref{asmp:irreducible}\ref{item:irreducible3}. It therefore follows from Lemma \ref{lem:feller} that, for all $n,n'\in\cN$, we have $\supp(\log G_{nn})\subset(2\pi/\tau)\Z$ and $\supp(\log G_{nn'})\subset a_{nn'}+(2\pi/\tau)\Z$ for some $a_{nn'}\in\R$. But this violates Assumption \ref{asmp:lattice}, so we conclude that $\mathrm{I}-\A(\alpha+i\tau)$ must be invertible for all nonzero $\tau\in\R$.
\end{proof}

\begin{proof}[Details of Example \ref{exmp:lattice}]
Observe that
\begin{equation*}
	\rho(\A(z))=\upsilon\rho\left(\begin{bmatrix}0&4^z\\2^z&0\end{bmatrix}\right)=2^{3z/2}\upsilon.
\end{equation*}
It follows that the unique positive solution to the equation $\rho(\A(z))=1$ is $z=-(2/3)\log_2\upsilon$, and so we deduce from Theorem \ref{thm:randomgrowth} that $S_t$ has a Pareto upper tail with exponent $\alpha=-(2/3)\log_2\upsilon$ in the sense of \eqref{eq:weakPareto}.

Assumption \ref{asmp:lattice} is not satisfied in this example. We now show that the conclusions of Theorem \ref{thm:randomgrowthlattice} do not hold. If reset occurs in period $\tau$ then, in the absence of further reset, the values taken by $S_{\tau+k}$ for $k\geq0$ are given by the (deterministic) sequence $1,4,8,32,64,256,512,\dots$. Therefore, if in period $t$ the most recent reset occurred in period $t-k$, then $S_t$ is equal to the $(k+1)$th entry in that sequence, which is $2^{\lfloor(3k+1)/2\rfloor}$. Since the probability that there have been exactly $k$ periods since the most recent reset\footnote{Here we implicitly extend $(S_t,J_t)$ to a stationary Markov process indexed by $t\in\Z$ rather than $t\in\Z_+$, so that the number of periods since the most recent reset is unbounded.} is $\upsilon^k(1-\upsilon)$, we find that
\begin{equation}\label{eq:latticeseries}
	\Pr(S_t>s)=\sum_{k=0}^\infty\upsilon^k(1-\upsilon)\mathbbm{1}(2^{\lfloor(3k+1)/2\rfloor}>s).
\end{equation}
Therefore, if $n-1\leq\log_2 s<n$, then
\begin{equation}\label{eq:latticetail1}
	\Pr(S_t>s)=\sum_{k=\lceil(2n-1)/3\rceil}^\infty\upsilon^k(1-\upsilon)=\upsilon^{\lceil(2n-1)/3\rceil}=\upsilon^{\lceil(2\lfloor\log_2s\rfloor+1)/3\rceil}.
\end{equation}
Multiplying by $s^\alpha$ with $\alpha=-(2/3)\log_2\upsilon$, we obtain
\begin{equation*}
	s^\alpha\Pr(S_t>s)=s^{-(2/3)\log_2\upsilon}\upsilon^{\lceil(2\lfloor\log_2s\rfloor+1)/3\rceil}=\upsilon^{-(2/3)\log_2s+\lceil(2\lfloor\log_2s\rfloor+1)/3\rceil},
\end{equation*}
which oscillates between $\upsilon$ and $\upsilon^{-1/3}$ as $s\to\infty$. Thus, while \eqref{eq:weakPareto} is satisfied, $s^\alpha\Pr(S_t>s)$ does not converge to any limit as $s\to\infty$.

The failure of Assumption \ref{asmp:lattice} also causes the convergence \eqref{eq:eigenvectorconvergencelattice} in Theorem \ref{thm:randomgrowthlattice} to fail in this example. To see why, observe first that, since the Markov state switches every period, summing the even-indexed terms in the series on the right-hand side of \eqref{eq:latticeseries} gives
\begin{equation*}
	\Pr(S_t>s,J_t=1)=\sum_{k=0}^\infty\upsilon^{2k}(1-\upsilon)\mathbbm{1}(2^{3k}>s).
\end{equation*}
Therefore, if $n-1\leq\log_2 s<n$, then
\begin{equation}\label{eq:latticetail2}
	\Pr(S_t>s,J_t=1)=\sum_{k=\lceil n/3\rceil}^\infty\upsilon^{2k}(1-\upsilon)=\frac{\upsilon^{2\lceil n/3\rceil}}{1+\upsilon}=\frac{\upsilon^{2\lceil(\lfloor\log_2s\rfloor+1)/3\rceil}}{1+\upsilon}.
\end{equation}
Dividing \eqref{eq:latticetail2} by \eqref{eq:latticetail1}, we obtain
\begin{equation*}
	\Pr(J_t=1\mid S_t>s)=\frac{\upsilon^{2\lceil(\lfloor\log_2s\rfloor+1)/3\rceil-\lceil(2\lfloor\log_2s\rfloor+1)/3\rceil}}{1+\upsilon},
\end{equation*}
which oscillates between $\upsilon/(1+\upsilon)$ and $1/(1+\upsilon)$ as $s\to\infty$. Thus $\Pr(J_t=1\mid S_t>s)$ does not converge to any limit as $s\to\infty$. Similarly, $\Pr(J_t=2\mid S_t>s)$ does not converge to any limit as $s\to\infty$.
\end{proof}

\section{Comparative statics}\label{sec:compstatappx}

We consider linear perturbations to the survival probabilities, to the parameters of location-scale transformations of the growth rates, and to a measure of the degree of persistence in the Markov modulator. Given $\mu_{nn'}\in \R$ and $\sigma_{nn'}>0$, we denote the MGF of $\mu_{nn'}+\sigma_{nn'}\log G_{nn'}$ by $\psi_{nn'}(z;\mu_{nn'},\sigma_{nn'})=\e^{z\mu_{nn'}}\psi_{nn'}(\sigma_{nn'}z)$, and normalize $\log G_{nn'}$ to have mean zero. Let $\Psi(z;\mathrm{M},\Sigma)$ be the $N\times N$ matrix of MGFs $\psi_{nn'}(z;\mu_{nn'},\sigma_{nn'})$ parametrized by $\mathrm{M}=(\mu_{nn'})$ and $\Sigma=(\sigma_{nn'})$. We parametrize persistence by setting $\Pi(\tau)=\tau \mathrm{I}+(1-\tau)\Pi$ for $\tau\in [0,1]$. Increases (decreases) in $\tau$ are interpreted as increases (decreases) in persistence.

Collect the parameters to be perturbed into a single vector of parameters,
\begin{equation*}
	\theta=(\Upsilon,\mathrm{M},\Sigma,\tau)\in (0,1)^{N\times N}\times \R^{N\times N}\times (0,\infty)^{N\times N}\times (0,1)\eqqcolon\Theta,
\end{equation*}
and let $\A(z;\theta)=\Pi(\tau)\odot\Upsilon\odot\Psi(z;\mathrm{M},\Sigma)$. The following result, an application of the implicit function theorem, shows how perturbations to $\theta$ affect $\alpha$.

\begin{prop}\label{thm:cs}
	Suppose the vector of parameters $\theta_0=(\Upsilon_0,\mathrm{M}_0,\Sigma_0,\tau_0)\in\Theta$ is such that $\Pi(\tau_0)$, $\Upsilon_0$ and $\Psi(z;\mathrm{M}_0,\Sigma_0)$ satisfy Assumption \ref{asmp:irreducible}, and the equation $\rho(\A(z;\theta_0))=1$ admits a unique positive solution $z=\alpha_0$ in the interior of
	\begin{align*}
		\cI_0\coloneqq\{z\in\R:\text{all entries of }\Psi(z;\mathrm{M}_0,\Sigma_0)\text{ are finite}\}.
	\end{align*}
	Then there exists a neighborhood $U\subset \Theta$ of $\theta_0$ and a unique continuously differentiable function $\alpha:U \to (0,\infty)$ such that $\rho(\A(\alpha(\theta);\theta))=1$ on $U$. The partial derivatives of $\alpha$ satisfy the following inequalities at $\theta_0$.
	\begin{enumerate}
		\item\label{item:cs.lifespan} $\partial\alpha/\partial {\upsilon_{nn'}}\le 0$: Longer lifespan implies a smaller Pareto exponent.	
		\item\label{item:cs.growth} $\partial\alpha/\partial\mu_{nn'}\le 0$: Higher growth implies a smaller Pareto exponent.
		\item\label{item:cs.volatility} $\partial\alpha/\partial\sigma_{nn'}\le 0$: Higher volatility implies a smaller Pareto exponent.
		\item\label{item:cs.persistence} If all columns or all rows of $\Upsilon_0\odot\Psi(z;\mathrm{M}_0,\Sigma_0)$ are the same, then $\partial\alpha/\partial\tau\le 0$: Higher persistence implies a smaller Pareto exponent.	
	\end{enumerate}
\end{prop}

\begin{rem}
	In Proposition \ref{thm:cs}\ref{item:cs.persistence}, the requirement that all columns or all rows of $\Upsilon_0\odot\Psi(z;\mathrm{M}_0,\Sigma_0)$ are the same means that the survival probability and growth rate distribution depend only on the current state, or only on the previous state. Without this condition, higher persistence need not imply a smaller Pareto exponent. As a counterexample, set $\cN=\set{1,2}$ and suppose that $G_{nn'}=2$ if $n\neq n'$ or $G_{nn'}=1$ if $n=n'$.	Suppose that the transition probability matrix for $J_t$ is
	\begin{equation*}
		\Pi(\tau)=\tau \mathrm{I}+(1-\tau)\begin{bmatrix}
			0 & 1 \\ 1 & 0
		\end{bmatrix}=\begin{bmatrix}
			\tau & 1-\tau \\ 1-\tau & \tau
		\end{bmatrix},
	\end{equation*}
	and the survival probability $\upsilon\in (0,1)$ is constant. Then
	\begin{equation*}
		\Pi(\tau)\odot\Upsilon\odot\Psi(z)=\upsilon\begin{bmatrix}
			\tau & (1-\tau)2^{z} \\
			(1-\tau)2^{z} & \tau
		\end{bmatrix},
	\end{equation*}
	whose spectral radius is $\upsilon(\tau+(1-\tau)2^{z})$. Setting the spectral radius equal to one and solving for $z>0$, we find that the Pareto exponent for the upper tail of the stationary distribution of $S_t$ is
	\begin{equation*}
		\alpha(\tau)=\log_2 \left(\frac{1/\upsilon-\tau}{1-\tau}\right)=\log_2 \left(1+\frac{1/\upsilon-1}{1-\tau}\right),
	\end{equation*}
	which is \emph{increasing} in $\tau\in (0,1)$. Therefore, in this example, increasing the persistence makes the upper tail lighter.
\end{rem}

\begin{proof}[Proof of Proposition \ref{thm:cs}]
	Define $F:(0,\infty)\times\Theta\to[-1,\infty]$ by
	\begin{align*}
		F(z;\theta)&=\rho(\A(z;\theta))-1,
	\end{align*}
	where the spectral radius is understood to be infinite whenever one or more entries of $\Psi(z;\mathrm{M},\Sigma)$ are infinite. Note that $F(\alpha_0;\theta_0)=0$ and that, since $\alpha_0$ lies in the interior of $\cI_0$, $F(z;\theta)$ is finite on a neighborhood of $(\alpha_0,\theta_0)$. To apply the implicit function theorem, we need $F$ to be continuously differentiable on a neighborhood of $(\alpha_0,\theta_0)$, with nonzero partial derivative $\partial F/\partial z$ at $(\alpha_0,\theta_0)$. Continuous differentiability follows from the fact that the entries of $\Pi(\tau)$ and $\Psi(z;\mathrm{M},\Sigma)$ are continuously differentiable with respect to their parameters and $z$, and the fact that the spectral radius of a nonnegative irreducible matrix is continuously differentiable with respect to its entries \citep[see e.g.][]{vahrenkamp1976}. Positivity of the partial derivative $\partial F/\partial z$ at $(\alpha_0,\theta_0)$ follows from Proposition \ref{prop:lambdaconv}. The implicit function theorem thus guarantees the existence of a neighborhood $U\subset\Theta$ of $\theta_0$ and a unique continuously differentiable function $\alpha:U\to (0,\infty)$ such that $\rho(\A(\alpha(\theta);\theta))=1$ on $U$. The partial derivatives of $\alpha$ on $U$ are then given by
	\begin{equation*}
		\nabla_\theta \alpha=-\frac{1}{\partial F/\partial z}\nabla_\theta F.
	\end{equation*}
	
	It remains to show that the partial derivatives of $\alpha$ have the signs asserted in \ref{item:cs.lifespan}--\ref{item:cs.persistence}. Since $\partial F/\partial z>0$, and since the spectral radius of a nonnegative matrix is nondecreasing in its entries \citep[Thm.~8.1.18]{HornJohnson2013}, to show \ref{item:cs.lifespan}--\ref{item:cs.volatility} it suffices to show that the entries of $\A(z;\theta)$ are nondecreasing in $\upsilon_{nn'}$, $\mu_{nn'}$ and $\sigma_{nn'}$. For $\upsilon_{nn'}$ this is obvious. For $\mu_{nn'}$ it follows from the fact that
	\begin{align*}
		\frac{\partial \psi_{nn'}(z;\mu_{nn'},\sigma_{nn'})}{\partial \mu_{nn'}}&=z\e^{z\mu_{nn'}}\psi_{nn'}(\sigma_{nn'}z),
	\end{align*}
	which is positive for $z>0$. For $\sigma_{nn'}$, observe that since $\log G_{nn'}$ is normalized to have zero mean, its MGF $\psi_{nn'}(z)$ satisfies $\psi_{nn'}'(0)=0$ and thus, by convexity, $\psi_{nn'}'(z)\geq0$ for $z>0$. The partial derivative
	\begin{align*}
		\frac{\partial \psi_{nn'}(z;\mu_{nn'},\sigma_{nn'})}{\partial \sigma_{nn'}}=z\e^{z\mu_{nn'}}\psi_{nn'}'(\sigma_{nn'}z)
	\end{align*}
    is therefore nonnegative for $z>0$, and so the entries of $\A(z;\theta)$ are nondecreasing in $\sigma_{nn'}$. To show \ref{item:cs.persistence}, observe that if all rows of $\Upsilon_0\odot\Psi(z;\mathrm{M}_0,\Sigma_0)$ are the same, then we may write $\A(z;\theta_0)=\Pi(\tau_0)\mathrm{D}_0(z)$, where $\mathrm{D}_0(z)=\diag(\Upsilon_0\odot\Psi(z;\mathrm{M}_0,\Sigma_0))$. On the other hand, if all columns of $\Upsilon_0\odot\Psi(z;\mathrm{M}_0,\Sigma_0)$ are the same, then we may write $\A(z;\theta_0)=\mathrm{D}_0(z)\Pi(\tau_0)$. Noting that $\rho(AB)=\rho(BA)$ in general \citep[Thm.~1.3.22]{HornJohnson2013}, in either case we have $\rho(\A(z;\theta_0))=\rho((\tau_0\mathrm{I}+(1-\tau_0)\Pi)\mathrm{D}_0(z))$. The fact that $\rho(\A(z;\theta_0))$ is nondecreasing in $\tau_0$ now follows from Theorem 5.2 in \citet{karlin1982}; see also Theorem 1.1 in \citet{altenberg2013}. Since $\partial F/\partial z>0$, it follows that $\alpha$ is nonincreasing in $\tau$ at $\theta_0$. This establishes \ref{item:cs.persistence}.
\end{proof}


\section*{References}
\begin{thebibliography}{99}
	
	\bibitem[Altenberg(2013)]{altenberg2013}
	Altenberg, Lee (2013): On the ordering of spectral radius product $r(\mathbf{A})r(\mathbf{AD})$ versus $r(\mathbf{A}^2\mathbf{D})$ and related applications, \emph{SIAM Journal on Matrix Analysis and Applications}, 34, 978--998. [\href{https://doi.org/10.1137/130906179}{DOI}]
	
	\bibitem[Beare, Seo and Toda(2022)]{BeareSeoToda2021}
	Beare, Brendan K., Won-Ki Seo, and Alexis A.\ Toda (2022): Tail behavior of stopped L\'{e}vy processes with Markov modulation, \emph{Econometric Theory}, in press. [\href{https://doi.org/10.1017/S0266466621000268}{DOI}]
	
	\bibitem[Beare and Toda(2020)]{BeareToda2020PhysD}
	Beare, Brendan K.\ and Alexis A.\ Toda (2020): On the emergence of a power law in the distribution of COVID-19 cases, \emph{Physica D: Nonlinear Phenomena}, 412, 132649. [\href{https://doi.org/10.1016/j.physd.2020.132649}{DOI}]
	
	\bibitem[Benhabib and Bisin(2018)]{benhabibbisin2018}
	Benhabib, Jess and Alberto Bisin (2018): Skewed wealth distributions: Theory and empirics, \emph{Journal of Economic Literature}, 56, 1261--1291. [\href{https://doi.org/10.1257/jel.20161390}{DOI}]
	
	\bibitem[Benhabib, Bisin and Zhu(2011)]{benhabib-bisin-zhu2011}
	Benhabib, Jess, Alberto Bisin, and Shenghao Zhu (2011): The distribution of wealth and fiscal policy in economies with finitely lived agents, \emph{Econometrica}, 79, 123--157. [\href{https://doi.org/10.3982/ECTA8416}{DOI}]
	
	\bibitem[Cao and Luo(2017)]{CaoLuo2017}
	Cao, Dan and Wenlan Luo (2017): Persistent heterogeneous returns and top end wealth inequality, \emph{Review of Economic Dynamics}, 26, 301--326. [\href{https://doi.org/10.1016/j.red.2017.10.001}{DOI}]
	
	\bibitem[Champernowne(1953)]{Champernowne1953}
	Champernowne, David G.\ (1953): A model of income distribution, \emph{Economic Journal}, 63, 318--351. [\href{https://doi.org/10.2307/2227127}{DOI}]
	
	\bibitem[Collamore(2009)]{collamore2009}
	Collamore, Jeffrey F.\ (2009): Random recurrence equations and ruin in a Markov-dependent stochastic economic environment, \emph{Annals of Applied Probability}, 19, 1404--1458. [\href{https://doi.org/10.1214/08-AAP584}{DOI}]
	
	\bibitem[Devadoss, Luckstead, Danforth and Akhundjanov(2016)]{DevadossLucksteadDanforthAkhundjanov2016}
	Devadoss, Stephen, Jeff Luckstead, Diana Danforth, and Sherzod Akhundjanov (2016): The power law distribution for lower tail cities in India, \emph{Physica A}, 442, 193--196. [\href{https://doi.org/10.1016/j.physa.2015.09.016}{DOI}]
	
	\bibitem[Douc, Moulines, Priouret and Soulier(2018)]{DoucMoulinesPriouretSoulier2018}
	Douc, Randal, Eric Moulines, Pierre Priouret, and Philippe Soulier (2018): \emph{Markov Chains}. Springer International Publishing.
	
	\bibitem[Embrechts, Kl\"{u}ppelberg and Mikosch(1997)]{EmbrechtsKluppelbergMikosch1997}
	Embrechts, Paul, Claudia Kl\"{u}ppelberg, and Thomas Mikosch (1997): \emph{Modelling Extremal Events: for Insurance and Finance}. Heidelberg: Springer-Verlag.
	
	\bibitem[Fagereng, Guiso, Malacrino and Pistaferri(2020)]{FagerengGuisoMalacrinoPistaferri2020}
	Fagereng, Andreas, Luigi Guiso, Davide Malacrino, and Luigi Pistaferri (2020): Heterogeneity and persistence in returns to wealth, \emph{Econometrica}, 88, 115--170.
	[\href{https://doi.org/10.3982/ECTA14835}{DOI}]
	
	\bibitem[Feller(1971)]{feller1971}
	Feller, William (1971): \emph{An Introduction to Probability Theory and its Applications, Vol.\ II.} New York: John Wiley \& Sons.
	
	\bibitem[Gabaix(1999)]{gabaix1999}
	Gabaix, Xavier (1999): Zipf's law for cities: An explanation, \emph{Quarterly Journal of Economics}, 114, 739--767. [\href{https://doi.org/10.1162/003355399556133}{DOI}]
	
	\bibitem[Gabaix(2009)]{gabaix2009}
	Gabaix, Xavier (2009): Power laws in economics and finance, \emph{Annual Review of Economics}, 1, 255--293. [\href{http://dx.doi.org/10.1146/annurev.economics.050708.142940}{DOI}]
	
	\bibitem[Gabaix and Ibragimov(2011)]{GabaixIbragimov2011}
	Gabaix, Xavier and Rustam Ibragimov (2011): Rank -- 1/2: A simple way to improve the OLS estimation of tail exponents, \emph{Journal of Business and Economic Statistics}, 29, 24--39. [\href{https://doi.org/10.1198/jbes.2009.06157}{DOI}]
	
	\bibitem[Gabaix(2016)]{gabaix2016}
	Gabaix, Xavier (2016): Power laws in economics: An introduction, \emph{Journal of Economic Perspectives}, 30, 185--206. [\href{https://doi.org/10.1257/jep.30.1.185}{DOI}]
	
	\bibitem[Gabaix, Lasry, Lions, and
	Moll(2016)]{GabaixLasryLionsMoll2016}
	Gabaix, Xavier, Jean-Michel Lasry, Pierre-Louis Lions, and Benjamin Moll (2016): The dynamics of inequality, \emph{Econometrica}, 84, 2071--2111. [\href{https://doi.org/10.3982/ECTA13569}{DOI}]
	
	\bibitem[Giesen, Zimmermann and Suedekum(2010)]{giesen-zimmermann-suedekum2010}
	Giesen, Kieran, Arndt Zimmermann, and Jens Suedekum (2010): The size distribution across all cities -- double Pareto lognormal strikes, \emph{Journal of Urban Economics}, 68, 129--137. [\href{https://doi.org/10.1016/j.jue.2010.03.007}{DOI}]
	
	\bibitem[Gomez and Gouin-Bonenfant(2020)]{GomezGouinbonenfant2020}
	Gomez, Matthieu and \'{E}milien Gouin-Bonenfant (2020): A Q-theory of inequality, Unpublished Manuscript, Columbia University. [\href{https://www.matthieugomez.com/files/qtheory.pdf}{URL}]
	
	\bibitem[Gouin-Bonenfant(2020)]{Gouinbonenfant2020}
	Gouin-Bonenfant, \'{E}milien (2020): Productivity dispersion, between-firm competition, and the labor share, Unpublished Manuscript, Columbia University. [\href{https://drive.google.com/open?id=1lNyuz2P1HXJKvfkM4FUMZxss4Tavl3Ai}{URL}]
	
	\bibitem[Gouin-Bonenfant and Toda(2022)]{Gouin-BonenfantTodaParetoExtrapolation}
	Gouin-Bonenfant, \'{E}milien and Alexis A.\ Toda (2022): Pareto extrapolation: An analytical framework for studying tail inequality, \emph{Quantitative Economics}, in press. [\href{https://papers.ssrn.com/sol3/papers.cfm?abstract_id=3260899}{URL}]
	
	\bibitem[Horn and Johnson(2013)]{HornJohnson2013}
	Horn, Roger A.\ and Charles R.\ Johnson (2013): \emph{Matrix Analysis}, 2nd ed. Cambridge, U.K.: Cambridge University Press.
	
	\bibitem[Karlin(1982)]{karlin1982}
	Karlin, Samuel (1982): Classifications of selection-migration structures and conditions for a protected polymorphism, in \emph{Evolutionary Biology, Vol.\ 14}, ed.\ by M.K.\ Hecht, B.\ Wallace and G.T.\ Prance. New York: Plenum, 61--204.
	
	\bibitem[Kesten(1973)]{kesten1973}
	Kesten, Harry (1973): Random difference equations and renewal theory for products of random matrices, \emph{Acta Mathematica}, 131, 207--248. [\href{https://doi.org/10.1007/BF02392040}{DOI}]
	
	\bibitem[Kingman(1961)]{Kingman1961}
	Kingman, J.F.C.\ (1961): A convexity property of positive matrices, \emph{Quarterly Journal of Mathematics}, 12, 283--284. [\href{https://doi.org/10.1093/qmath/12.1.283}{DOI}]
	
	\bibitem[Ma, Stachurski and Toda(2020)]{MaStachurskiToda2020}
	Ma, Qingyin, John Stachurski, and Alexis A.\ Toda (2020): The income fluctuation problem and the evolution of wealth, \emph{Journal of Economic Theory}, 187, 105003. [\href{https://doi.org/10.1016/j.jet.2020.105003}{DOI}]
	
	\bibitem[Manrubia and Zanette(1999)]{ManrubiaZanette1999}
	Manrubia, Susanna C.\ and Dami\'{a}n H.\ Zanette (1999): Stochastic multiplicative processes with reset events, \emph{Physical Review E}, 59, 4945--4948. [\href{https://doi.org/10.1103/PhysRevE.59.4945}{DOI}]
	
	\bibitem[Meylahn, Sabhapandit and Touchette(2015)]{MeylahnSabhapanditTouchette2015}
	Meylahn, Janusz M., Sanjib Sabhapandit, and Hugo Touchette (2015): Large deviations for Markov processes with resetting, \emph{Physical Review E}, 92, 062148. [\href{https://doi.org/10.1103/PhysRevE.92.062148}{DOI}]
	
	\bibitem[Mitzenmacher(2004)]{Mitzenmacher2004}
	Mitzenmacher, Michael (2004): A brief history of generative models for power law and lognormal distributions, \emph{Internet Mathematics}, 1, 226--252. [\href{https://doi.org/10.1080/15427951.2004.10129088}{DOI}]
	
	\bibitem[Montero and Villarroel(2013)]{MonteroVillarroel2013}
	Montero, Miquel and Javier Villarroel (2013): Monotonic continuous-time random walks with drift and stochastic reset events, \emph{Physical Review E}, 87, 012116. [\href{https://doi.org/10.1103/PhysRevE.87.012116}{DOI}]
	
	\bibitem[Mukoyama and Osotimehin(2019)]{MukoyamaOsotimehin2019}
	Mukoyama, Toshihiko and Sophie Osotimehin (2019): Barriers to reallocation and economic growth: The effects of firing costs, \emph{American Economic Journal: Macroeconomics}, 11, 235--270. [\href{https://doi.org/10.1257/mac.20170170}{DOI}]
	
	\bibitem[Nakagawa(2007)]{Nakagawa2007}
	Nakagawa, Kenji (2007): Application of Tauberian theorem to the exponential decay of the tail probability of a random variable, \emph{IEEE Transactions on Information Theory}, 53, 3239--3249. [\href{https://doi.org/10.1109/TIT.2007.903114}{DOI}]
	
	\bibitem[Nirei and Aoki(2016)]{NireiAoki2016}
	Nirei, Makoto and Shuhei Aoki (2016): Pareto distribution of income in neoclassical growth models, \emph{Review of Economic Dynamics}, 20, 25--42. [\href{https://doi.org/10.1016/j.red.2015.11.002}{DOI}]
	
	\bibitem[Reed(2001)]{reed2001}
	Reed, William J.\ (2001): The Pareto, Zipf and other power laws, \emph{Economics Letters}, 74, 15--19. [\href{https://doi.org/10.1016/S0165-1765(01)00524-9}{DOI}]
	
	\bibitem[Reed(2002)]{reed2002}
	Reed, William J.\ (2002): On the rank-size distribution for human settlements, \emph{Journal of Regional Science}, 42, 1--17. [\href{https://doi.org/10.1111/1467-9787.00247}{DOI}]
	
	\bibitem[Roitershtein(2007)]{roitershtein2007}
	Roitershtein, Alexander (2007): One-dimensional linear recursions with Markov-dependent coefficients, \emph{Annals of Applied Probability}, 17, 572--608. [\href{https://doi.org/10.1214/105051606000000844}{DOI}]
	
	\bibitem[de Saporta(2005)]{desaporta2005}
	de Saporta, Beno\^{i}te (2005): Tail of the stationary solution of the stochastic equation $Y_{n+1}=a_nY_n+b_n$ with Markovian coefficients, \emph{Stochastic Processes and their Applications}, 115, 519--578. [\href{https://doi.org/10.1016/j.spa.2005.06.009}{DOI}]
	
	\bibitem[Simon and Bonini(1958)]{SimonBonini1958}
	Simon, Herbert A.\ and Charles P.\ Bonini (1958): The size distribution of business firms, \emph{American Economic Review}, 48, 607--617. [\href{https://doi.org/10.2307/3003596}{DOI}]
	
	\bibitem[Toda(2011)]{Toda2011PRE}
	Toda, Alexis A.\ (2011): Income dynamics with a stationary double Pareto distribution, \emph{Physical Review E}, 83, 046122. [\href{https://doi.org/10.1103/PhysRevE.83.046122}{DOI}]
	
	\bibitem[Toda(2012)]{Toda2012JEBO}
	Toda, Alexis A.\ (2012): The double power law in income distribution: Explanations and evidence, \emph{Journal of Economic Behavior and Organization}, 84, 364--381. [\href{https://doi.org/10.1016/j.jebo.2012.04.012}{DOI}]
	
	\bibitem[Toda(2014)]{Toda2014JET}
	Toda, Alexis A.\ (2014): Incomplete market dynamics and cross-sectional distributions, \emph{Journal of Economic Theory}, 154, 310--348. [\href{https://doi.org/10.1016/j.jet.2014.09.015}{DOI}]
	
	\bibitem[Toda(2017)]{Toda2017MD}
	Toda, Alexis A.\ (2017): A note on the size distribution of consumption: More double Pareto than lognormal, \emph{Macroeconomic Dynamics}, 21, 1508--1518. [\href{https://doi.org/10.1017/S1365100515000942}{DOI}]
	
	\bibitem[Toda(2019)]{Toda2019}
	Toda, Alexis A.\ (2019): Wealth distribution with random discount factors, \emph{Journal of Monetary Economics}, 104, 101--113. [\href{https://doi.org/10.1016/j.jmoneco.2018.09.006}{DOI}]
	
	\bibitem[Toda and Walsh(2015)]{TodaWalsh2015JPE}
	Toda, Alexis A.\ and Kieran Walsh (2015): The double power law in consumption and implications for testing Euler equations, \emph{Journal of Political Economy}, 123, 1177--1200. [\href{https://doi.org/10.1086/682729}{DOI}]
	
	\bibitem[Vahrenkamp(1976)]{vahrenkamp1976}
	Vahrenkamp, Richard (1976): Derivatives of the dominant root, \emph{Applied Mathematics and Computation}, 2, 29--39. [\href{https://doi.org/10.1016/0096-3003(76)90018-7}{DOI}]
	
	\bibitem[Wold and Whittle(1957)]{WoldWhittle1957}
	Wold, Herman O.A.\ and Peter Whittle (1957): A model explaining the Pareto distribution of wealth, \emph{Econometrica}, 25, 591--595. [\href{https://doi.org/10.2307/1905385}{DOI}]
	
	\bibitem[Yamamoto(2014)]{Yamamoto2014}
	Yamamoto, Ken (2014): Stochastic model of Zipf's law and the universality of the power-law exponent, \emph{Physical Review E}, 89, 042115. [\href{https://doi.org/10.1103/PhysRevE.89.042115}{DOI}]
	
\end{thebibliography}
\end{document}